%% file: sample-manuscript.tex
\documentclass[manuscript, screen]{acmart}

\AtBeginDocument{%
  }

\setcopyright{acmlicensed}
\acmYear{2025}
\acmMonth{10}
\acmJournal{TQC}
\acmDOI{}
\acmVolume{}
\acmNumber{}
\acmArticle{}

\usepackage{algorithm}
\usepackage[noend]{algpseudocode}

\begin{document}

\title[QASMTrans: An End-to-End QASM Compilation Framework with Pulse Generation]{QASMTrans: An End-to-End QASM Compilation Framework with Pulse Generation for Near-Term Quantum Devices}

\author{Aaron Hoyt}
\orcid{0000-0001-9877-3236}
\affiliation{
    \institution{University of Washington}
    \city{Seattle}
    \country{USA}
}
\affiliation{
    \institution{Pacific Northwest National Laboratory}
    \city{Richland}
    \country{USA}
}

\author{Meng Wang}
\affiliation{
    \institution{Pacific Northwest National Laboratory}
    \city{Richland}
    \country{USA}
}

\author{Fei Hua}
\affiliation{
    \institution{Pacific Northwest National Laboratory}
    \city{Richland}
    \country{USA}
}

\author{Chunshu Wu}
\affiliation{
    \institution{Pacific Northwest National Laboratory}
    \city{Richland}
    \country{USA}
}

\author{Chenxu Liu}
\affiliation{
    \institution{Pacific Northwest National Laboratory}
    \city{Richland}
    \country{USA}
}

\author{Muqing Zheng}
\affiliation{
    \institution{Pacific Northwest National Laboratory}
    \city{Richland}
    \country{USA}
}

\author{Samuel Stein}
\affiliation{
    \institution{Pacific Northwest National Laboratory}
    \city{Richland}
    \country{USA}
}

\author{Drew Rebar}
\affiliation{
    \institution{Pacific Northwest National Laboratory}
    \city{Richland}
    \country{USA}
}

\author{Yufei Ding}
\affiliation{
    \institution{University of California San Diego}
    \city{San Diego}
    \country{USA}
}

\author{Travis S. Humble}
\affiliation{
    \institution{Oak Ridge National Laboratory}
    \city{Oak Ridge}
    \country{USA}
}

\author{Ang Li}
\affiliation{
    \institution{Pacific Northwest National Laboratory}
    \city{Richland}
    \country{USA}
}
\affiliation{
    \institution{University of Washington}
    \city{Seattle}
    \country{USA}
}


\begin{abstract}
 
QASMTrans is a lightweight, high-performance, C++-based quantum compiler that bridges abstract quantum algorithms to device-level control and is designed for just-in-time (JIT) deployment on QPU testbeds with tightly integrated FPGAs or CPUs. We focus on achieving fast transpilation times on circuits of interest, we find more than 100× faster compilation than Qiskit in some circuits with similar circuit quality, enabling transpilation of large, high-depth circuits in seconds. Unlike existing tools, QASMTrans offers end-to-end device-pulse compilation and direct quantum control integration with QICK, closing the gap between logical circuits and hardware control enabling closed-loop optimization. QASMTrans supports latency-aware Application-tailored Gate Sets (AGS) at the pulse level, identifying high-impact gate sequences on the circuit critical path and synthesizing optimized pulse schedules using pre-defined robust circuit ansatz. Validated through integrated QuTiP pulse-level simulation, this is found to significantly reduce execution latency and can improve final-state fidelity by up to 12\% in some tested circuits. QASMTrans further implements device-aware, noise-adaptive transpilation that uses device calibration data for circuit placement on high-quality qubits and can focus on the circuit critical path to reduce transpilation-pass time while maintaining comparable fidelity. Additionally, it introduces circuit space sharing via calibration-aware device partitioning, enabling concurrent execution of multiple circuits or shots on a single QPU. Moreover, QASMTrans is entirely self-contained and has no external library dependencies, making it easy for practical deployment. We validate QASMTrans across IBM, Rigetti, IonQ, and Quantinuum platforms, demonstrating $<$1\% fidelity deviation from Qiskit while delivering consistent performance from ARM-based embedded devices to leadership-class HPC systems. By combining fast compilation, pulse-level control, and noise-aware optimization, QASMTrans enables real-time adaptive algorithms such as ADAPT-VQE and ADAPT-QAOA. Source code: http://github.com/pnnl/qasmtrans

\end{abstract}



\received{30 October 2025}

\maketitle
\renewcommand{\shortauthors}{Hoyt et al.}

\input{introduction}
\input{background}
\input{Transpiler}

\input{Pulses}
\input{Adaptive}

\input{Python}
\input{evaluation}
\input{relatedwork}

\section{Conclusion} \label{sec:conclude}

We presented QASMTrans, a C++-based, high-performance quantum compiler designed for deployment in QPU control systems for JIT compilation and in HPC environments for large-scale circuit processing. QASMTrans addresses several unsolved problems in the quantum software stack. First, it provides end-to-end device pulse control, enabling compilation from abstract quantum programs to calibrated pulse schedules through device configuration files. QASMTrans is the first open-source compiler to fully support the quantum stack from logical-level programs to pulse control via QICK integration. Second, the framework demonstrates application-aware pulse optimization through latency-aware gate merging that targets the circuit critical path, reducing execution latency by up to 31\% and improving fidelity by up to 12\%. Third, QASMTrans implements noise-adaptive transpilation with critical-path-aware placement heuristics and introduces compiler-supported space sharing for concurrent circuit execution on partitioned devices.

The framework's minimal dependencies and efficient C++ implementation enable deployment across diverse platforms from ARM-based embedded devices to leadership-class HPC systems, achieving up to 171× speedup over Qiskit . A Python API provides programmatic access for seamless workflow integration. 

QASMTrans is released as open-source software at to serve as both a foundation for quantum compilation research and a practical tool for testbed operators, enabling the community to efficiently develop, optimize, and execute increasingly complex quantum algorithms on near-term NISQ devices. \url{http://github.com/pnnl/qasmtrans}

\section{Future Work} \label{sec:future}

Future work will focus on expanding front-end optimization passes including advanced gate cancellation. Device support will be extended to distributed quantum systems and emerging cavity-based architectures, with additional intermediate representations such as QIR to broaden interoperability. Hardware validation of AGS pulses through established calibration pipelines (DRAG, RB/ORBIT, simultaneous RB) will close the loop between simulation and physical execution. We will also integrate QASMTrans with NWQ-Control package to enable this closed loop control optimization with the physical hardware. Finally, integration with commercial QPU control systems such as those from IQM will validate QASMTrans for production quantum testbed deployment.

\begin{acks}
This material is mainly based upon work supported by the U.S. Department of Energy, Office of Science, National Quantum Information Science Research Centers, Quantum Science Center (QSC). Some early-stage developments of the original version of QASMTrans were supported by the U.S. Department of Energy, Office of Science, National Quantum Information Science Research Centers, Co-design Center for Quantum Advantage (C2QA) under contract number DE-SC0012704. This research used resources of the Oak Ridge Leadership Computing Facility, which is a DOE Office of Science User Facility supported under Contract DE-AC05-00OR22725. This research used resources of the National Energy Research Scientific Computing Center (NERSC), a U.S. Department of Energy Office of Science User Facility located at Lawrence Berkeley National Laboratory, operated under Contract No. DE-AC02-05CH11231.
The Pacific Northwest National Laboratory is operated by Battelle for the U.S. Department of Energy under Contract DE-AC05-76RL01830.
\end{acks}

\bibliographystyle{ACM-Reference-Format}
\bibliography{references, refs}

\end{document}

%% file: introduction.tex
\section{Introduction}

The past decade has witnessed tremendous development in \emph{Noisy Intermediate-Scale Quantum} (NISQ) computers \cite{clarke2008superconducting, rigetti2012superconducting}, where a few hundred physical qubits are available with relatively limited coherence times and high error rates. These NISQ machines, while offering great potential, are constrained by various factors such as non-trivial noise~\cite{maciejewski2020mitigation}\cite{tannu+:micro19}, limited connectivity~\cite{chamberland2020topological} and machine-specific basis gate sets~\cite{lin2022let}. Due to the limited qubit number and short coherence time, effectively mapping application circuits to the constrained NISQ machine poses a considerable challenge and can significantly impact the fidelity of the execution results.
Transpilation is the specific terminology referring to the compilation process of transforming a high-level quantum circuit into an equivalent circuit that is compatible with the specifications of a quantum device, including: the basis gate set, topology of the quantum chip, timing constraints, fidelity of operations, etc. The goal of a transpiler is to perform this transformation while minimizing the impact on the functionality of the circuit and optimizing its performance delivery.

Several attempts on quantum transpilation have already been made by the community (see a summary in Section~\ref{sec:related_work}), but there are still technical gaps. On the one hand, commercial transpilers such as those embedded in Qiskit~\cite{IBMQiskit} and Cirq~\cite{cirq} provide comprehensive functionalities, but are typically slow, especially for deep circuits arising in practical quantum applications such as chemistry~\cite{kauzmann+:qcbook13, cao+:cr19}, optimization~\cite{dunjko+:rpp18, wichert2020principles} and nuclear physics~\cite{stetcu2022projection, holmes2022quantum}. Additionally, the slow transpilation speed limits their capability to explore a larger design space and integrate more advanced but expensive optimizations. This is especially the case when dynamic circuit generation and transpilation is needed, such as in variational quantum algorithms (VQAs)~\cite{cerezo2021variational, stein2022eqc} and when optimized to mitigate state-dependent bias at runtime \cite{tannu2019mitigating}.

On the other hand, most of the research studies in academia have focused on specific transpilation techniques, such as gate decomposition, circuit optimization, mapping and routing, etc.~\cite{li+:asplos19}\cite{zulehner+:date18:}\cite{zhang+:asplos21}. These approaches lack end-to-end demonstrations and are often implemented and validated by embedding into or replacing part of Python-based commercial frameworks such as Qiskit and Cirq. Consequently, they are also constrained by the limitations of the underlying frameworks, such as slow speed, difficulties in launching large circuits, binding to certain device features, lack of flexibility, and frequent interface upgrades, etc. 

Beyond gate-level compilation, emerging quantum testbeds require pulse-level control to enable fine-grained optimization and adaptive algorithm execution. However, existing transpilers provide limited support for embedded systems with direct QPU integration. To date, only Qiskit and XACC \cite{9347736, alexander+:qiskitpulse20} offer pulse-level transpilation \cite{Cross_2022}. Moreover, Qiskit restricts pulse operations to IBM hardware, while XACC only supports pulse compilation to its pulse simulator. This gap is critical for variational quantum algorithms such as ADAPT-VQE, which dynamically constructs ansatze based on measurement feedback \cite{Grimsley_2019}, and variants like ADAPT-QAOA that grow circuits after each evaluation \cite{PhysRevA.109.032420,zhu2022adaptivequantumapproximateoptimization}. These algorithms require real-time transpilation with minimal overhead—a capability not addressed by current frameworks.

Recent work has demonstrated the potential of pulse-level optimization. Wei et al. showed the ability to build arbitrary native two qubit pulse gates on IBM systems \cite{PRXQuantum.5.020338}, while Chen et al. comprehensively studied mining transpiled circuits for frequent sub-circuits and merging gates to decrease latency \cite{chen2023pulse}. Recently the AshN gate pulse ansatz has been demonstrated to realize arbitrary two qubit unitaries on a superconducting quantum computer from Chen et. al. \cite{chen2025efficient}.  Building on these insights, we take a complementary approach by targeting the critical path of quantum circuits and designing customized pulse gates for frequent pre-transpiled gates. This strategy reduces the overhead of finding circuits to calibrate during transpilation while enabling substantial reductions in circuit latency and improvements in fidelity.

\begin{figure}[htb]
    \centering
    \begin{minipage}[c]{0.49\linewidth}
        \centering
        \includegraphics[width=\linewidth]{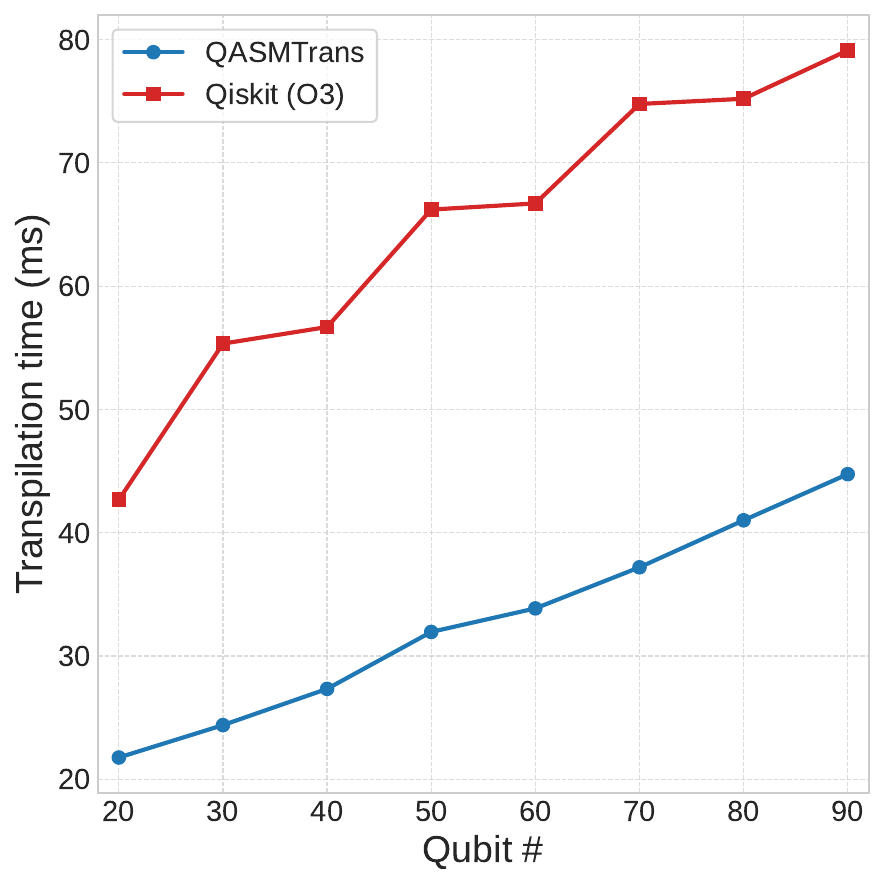}
    \end{minipage}\hfill
    \begin{minipage}[c]{0.49\linewidth}
        \centering
        \includegraphics[width=\linewidth]{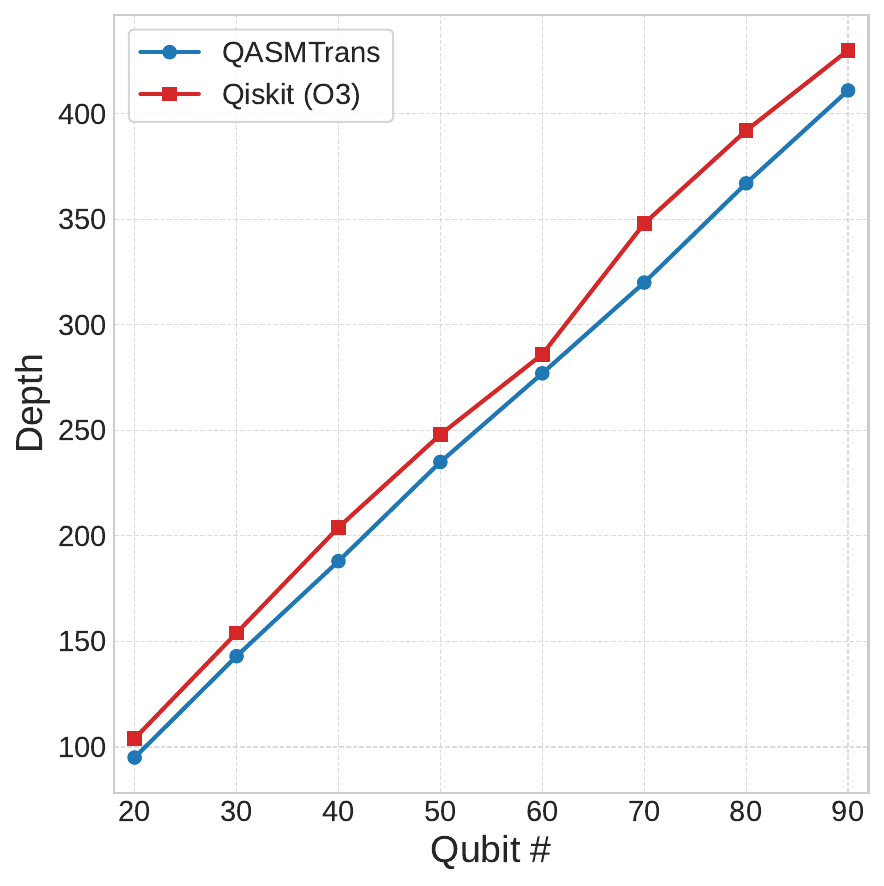}
    \end{minipage}

    \vspace{2pt}
    \makebox[0.49\linewidth][c]{\textbf{a)}}\hfill
    \makebox[0.49\linewidth][c]{\textbf{b)}}

    \caption{QASMTrans is designed to transpile challenging deep circuits. Performance comparison on VQE-UCCSD circuits: (a) QASMTrans achieves nearly a twofold reduction in compilation time relative to Qiskit O3, and (b) produces comparable circuit depth.}
    \label{fig:compilation_time}
\end{figure}

In this paper, we present QASMTrans, an end-to-end, self-contained, light-weight quantum transpiler entirely realized in C++ for effectively parsing and compiling large QASM circuits. QASMTrans integrates pulse-level compilation with direct QICK interface for instant and efficient QPU control through transpilation, validated through a custom QuTiP-based simulator modeling the 9-qubit Rigetti Ankaa chip ~\cite{PhysRevApplied.16.024050}. QASMTrans comprises four major components: 
\begin{enumerate}
    \item An {\textbf{\emph{IO}}} module that uses a QASM Parser for parsing an input OpenQASM file, and translating it into a structure acting as the internal intermediate representation (IR). The output will export the transpiled QASM circuits for a particular NISQ device, such as those provided by IBMQ, Rigetti, IonQ, Quantinuum, etc.
    \item A {\textbf{\emph{Configuration}}} module for preparing the coupling graph of the device, generating the DAG for the circuit, and decomposing the 3-qubit gates into 1-qubit and 2-qubit gates.
    \item An {\textbf{\emph{Optimization}}} module for the various optimization passes. This includes decomposition into basis gates, routing, and mapping. These passes are made with respect to the topology, basis gate set, fidelity, and features of the circuit. The goal of the backend optimization is to allow the circuits to run more efficiently on the targeted NISQ devices or simulators.
    \item A main \textbf{\emph{Transpiler}} component to do the routing and mapping and also decompose into basis gates based on specific NISQ devices.
\end{enumerate}

QASMTrans is primarily designed as an open-source transpiler infrastructure serving as a baseline for implementing and validating advanced transpilation technologies while supporting novel devices and computation models. In Fig. \ref{fig:compilation_time} we demonstrate performance scaling against Qiskit as well as comparative quality against the VQE-UCCSD algorithm. We evaluate QASMTrans using diverging circuits with some of them being quite challenging (from 4 to 127 qubits, and from 10 to 500K gates, see Figure~\ref{fig:compilation_time}) from QASMBench~\cite{li2023qasmbench}. We find most of the benchmarks can be completed within a few seconds. In all transpilation cases QASMTrans can beat Qiskit in transpilation time efficiency. This work thus makes the following main contributions:
\begin{itemize}
    \item We propose an end-to-end, self-contained, light-weighted opensource quantum compiler in C++ that can significantly reduce the transpilation time for a wide range of applications, improving the efficiency of quantum computations on NISQ devices.
    \item QASMTrans is equipped with optimization techniques for generating specific basis gates towards different target machines or classical simulators. 
    \item Through comprehensive experiments and analysis over multiple quantum platforms, we show that QASMTrans can transpile circuits with comparable fidelity on real NISQ devices from Rigetti, IBMQ, IonQ, and Quantinuum, but at a much faster speed compared to existing transpilers such as Qiskit.  
\end{itemize}
The remainder of this paper is structured as follows: Section~\ref{sec:background} provides background information. Section~\ref{sec:transpiler} presents the QASMTrans transpiler.
Section~\ref{sec:pulse} presents pulse compilation.
Section~\ref{sec:adaptive} presents noise adaptive compilation and circuit sharing.
Section~\ref{sec:python} discusses the python API.
Section~\ref{sec:evaluation} shows the evaluation results. Section~\ref{sec:related_work} summarizes related work about quantum transpilation. Section~\ref{sec:conclude} concludes.

%% file: background.tex
\section{Background} \label{sec:background}

\subsection{Noisy Intermediate-Scale Quantum (NISQ)}

NISQ systems refer to near-term quantum platforms featuring tens to thousands of qubits \cite{preskill2018quantum}. These qubits are based on various technologies, such as superconducting \cite{clarke2008superconducting, rigetti2012superconducting}, trapped-ion \cite{cirac1995quantum, leibfried2003quantum}, photonic \cite{o2009photonic, aspuru2012photonic}, spin qubits \cite{pla2012single, maurand2016cmos}, neutral atoms \cite{briegel2000quantum, henriet2020quantum}, etc. To accomplish the execution of a circuit, the physical qubits need to stay coherent for a sufficiently long time. However, before all the circuits can be executed on the real quantum machine, it must (1) fit the basis gates of the quantum machine and (2) meet the coupling constraints of the machine topology.

\subsubsection{Basis Gates}
\begin{table}[h!]
\centering\normalsize
\caption{Basis gates for IBM-Q, Rigetti, IonQ, and Quantinuum NISQ devices.}
\begin{tabular}{|c|c|c|c|}
\hline
\textbf{NISQ} & \textbf{Technology} & \textbf{1-qubit basis} & \textbf{2-qubit basis}   \\ \hline
IBMQ & Superconducting & \texttt{ID}, \texttt{RZ}, \texttt{SX}, \texttt{X} & \texttt{CX/ECR} \\ \hline
Rigetti & Superconducting & \texttt{RX}, \texttt{RZ} & \texttt{CZ} (\texttt{XY})  \\ \hline
IonQ & Trapped-Ion & \texttt{GPI}, \texttt{GPI2}, \texttt{GZ} & \texttt{MS} \\ \hline
Quantinuum & Trapped-Ion & \texttt{RX}, \texttt{RZ} & \texttt{ZZ}  \\ \hline
\end{tabular}
\label{tab:basis_gates}
\end{table}

\begin{figure}[htb]
    \centering
    \begin{minipage}[c]{0.24\linewidth}
        \centering
        \includegraphics[width=\linewidth]{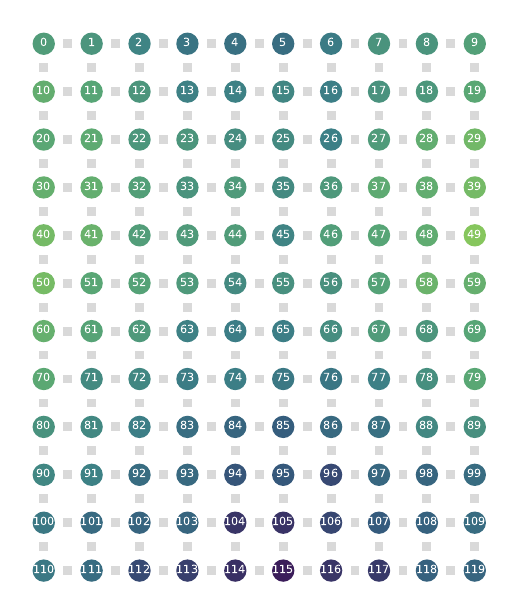}
    \end{minipage}\hfill
    \begin{minipage}[c]{0.24\linewidth}
        \centering
        \includegraphics[width=\linewidth]{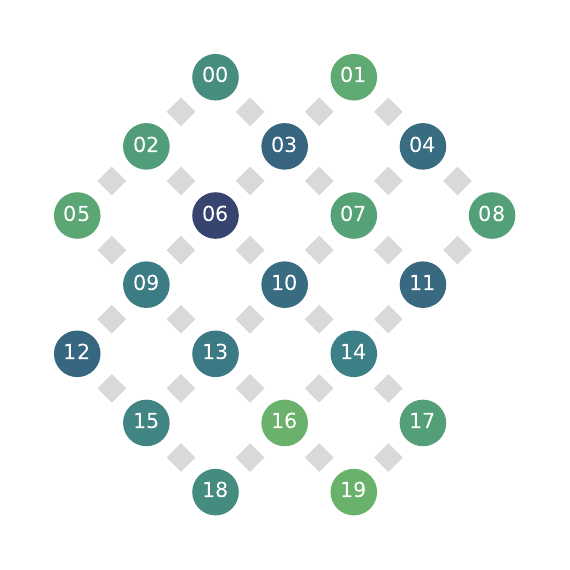}
    \end{minipage}
        \begin{minipage}[c]{0.24\linewidth}
        \centering
        \includegraphics[width=\linewidth]{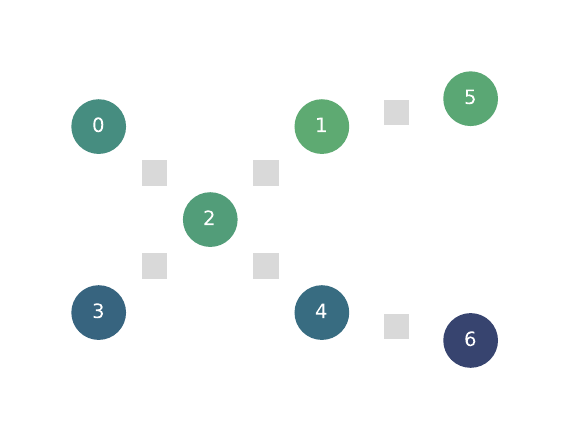}
    \end{minipage}
        \begin{minipage}[c]{0.24\linewidth}
        \centering
        \includegraphics[width=\linewidth]{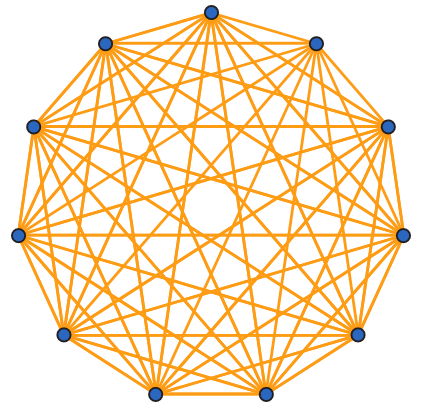}
    \end{minipage}

    \vspace{2pt}
    \makebox[0.25\linewidth][c]{\textbf{a)}}\hfill
    \makebox[0.25\linewidth][c]
    {\textbf{b)}}\hfill
    \makebox[0.25\linewidth][c]{\textbf{c)}}\hfill
    \makebox[0.25\linewidth][c]{\textbf{d)}}

    \caption{NISQ device topology: (a) IBMQ-Miami, (b) IQM Crystal 20, (c) Quantware Soprano D5, (d) IonQ-QPU}
    \label{fig:ibmq_topo}
\end{figure}

Each NISQ device has its own basis gate set, known as the \emph{quantum instruction set architecture} (QISA). It defines the basic operations that are physically supported by the underlying platform. During quantum transpilation, all the logic gates will be decomposed and transpiled into gate sequences purely formed by basis gates.  Table~\ref{tab:basis_gates} shows the basis gate set for IBMQ, Rigetti, IonQ, and Quantinuum devices. Typically, quantum device vendors only provide profiling or calibration data for the basis gates (per qubit or system-wide average), including T1, T2, duration, fidelity, etc. These basis gates also represent the operations to be implemented by a classical simulator.
\subsubsection{Topology}

Physical qubits in a quantum processor are interconnected. In a quantum device, the 1-qubit gates are directly performed on individual qubits. The 2-qubit gates, however, have to be performed on a qubit-pair that is interconnected. This is especially the case for superconducting devices (e.g., IBMQ and Rigetti), where the connectivity of qubits follows a certain topology, as shown in Figure~\ref{fig:ibmq_topo}. The topology thus limits the sites where two-qubit gates can be performed: if a two-qubit gate is desired for remote qubits, a series of \texttt{SWAP} gates are required to move the quantum information from two qubits to a connected tuple following the path defined by the topology, known as \emph{routing}. \texttt{SWAP} gates are costly, usually achieved through three \texttt{CNOT} or \texttt{CX} gates. 

These extra \texttt{SWAP}s are one of the major factors contributing to deep circuits and considerable noise for superconducting devices, as compared to contemporary small-scale trapped-ion devices practicing all-to-all connectivity (see Figure~\ref{fig:ibmq_topo}). Our previous study \cite{stein2022quclassi} shows that, for a 17-gate variational circuit, from the 5-qubit IBMQ Cairo to the 5-qubit IonQ QPU, a fidelity increase from 72\% to 80\% (ideally 97.8\%) has been observed. This is mainly due to the 7 extra \texttt{SWAP} gates when transpiling to comply with the topology of IBMQ Cairo.

\subsection{QASM}

\begin{table*}[h!]
\centering\normalsize
\caption{OpenQASM gate definition (5 basic gates + 11 standard gates + 18 composition gates).}
\begin{tabular}{|c|l|c|l|c|l|}
\hline
\textbf{Gates} & \textbf{Meaning} & \textbf{Gates} & \textbf{Meaning} & \textbf{Gates} & \textbf{Meaning}  \\ \hline
\texttt{U3} & 3 parameter 2 pulse 1-qubit & \texttt{TDG} & conjugate of sqrt(S) &  \texttt{CRZ} & Controlled RZ rotation  \\ \hline
\texttt{U2} & 2 parameter 1 pulse 1-qubit  & \texttt{RX} & X-axis rotation    &  \texttt{CU1} & Controlled phase rotation   \\ \hline
\texttt{U1} & 1 parameter 0 pulse 1-qubit  & \texttt{RY} & Y-axis rotation  &  \texttt{CU3} & Controlled U3   \\ \hline
\texttt{CX} & Controlled-NOT & \texttt{RZ} & Z-axis rotation     &  \texttt{RXX} & 2-qubit XX rotation   \\ \hline
\texttt{ID} & Idle gate or identity & \texttt{CZ} & Controlled phase    & \texttt{RZZ} & 2-qubit ZZ rotation   \\ \hline
\texttt{X} & Pauli-X bit flip   & \texttt{CY} & Controlled Y   & \texttt{RCCX} & Relative-phase CXX    \\ \hline
\texttt{Y} & Pauli-Y bit and phase flip   & \texttt{SWAP} & Swap   & \texttt{RC3X} & Relative-phase 3-controlled X    \\ \hline
\texttt{Z} & Pauli-Z phase flip   &  \texttt{CH} & Controlled H   & \texttt{C3X} & 3-controlled X  \\ \hline
\texttt{H} & Hadamard   &  \texttt{CCX} & Toffoli  & \texttt{C3XSQRTX} & 3-controlled sqrt(X)   \\ \hline
\texttt{S} & sqrt(Z) phase  &  \texttt{CSWAP} & Fredkin   & \texttt{C4X} & 4-controlled X  \\ \hline
\texttt{SDG} & conjugate of sqrt(Z)  & \texttt{CRX} & Controlled RX rotation & &   \\ \hline
\texttt{T} & sqrt(S) phase   & \texttt{CRY} & Controlled RY rotation & & \\ \hline
\end{tabular}
\label{tab:gates}
\end{table*}
OpenQASM (Open Quantum Assembly Language, we particularly refer to OpenQASM~2.0 in this work) \cite{cross2017open}, also known colloquially as QASM, is an intermediate representation (IR) of quantum instructions. QASM acts as a unified low-level assembly language for IBMQ and other quantum machines. Many of these NISQ devices, accessible through the IBMQ network \cite{ibm}, have been widely explored by existing works. 
Table~\ref{tab:gates} lists the types of gates that are defined in the QASM specification (i.e., the "\texttt{qelib1.inc}" header file) \cite{cross2017open}. Within these gates, the first five, i.e., \texttt{U3}, \texttt{U2}, \texttt{U1}, \texttt{CX}, and \texttt{ID}, are \emph{basic gates} that are expected to be supported by the quantum backend. From \texttt{X} to \texttt{RZ} are \emph{standard gates} defined atomically in OpenQASM. The remaining gates from \texttt{CZ} to \texttt{C4X} are \emph{composition gates} that are constructed by standard gates. These gates are frequently used gates defined in \texttt{qelib1.inc}.
OpenQASM~2.0 is a low-level IR, which is executed sequentially without any loops, branches, or jumps, making it very convenient for static analysis and simulating in a classical simulator~\cite{li2020density, li2021sv}. A QASM code can be directly launched in IBMQ or through Qiskit. With all these benefits, QASMTrans uses QASM as the primary format for input and output.

%% file: Transpiler.tex
\section{QASMTrans Transpiler} 

\begin{figure}[h]
  \centering
  \includegraphics[width=0.7\linewidth]{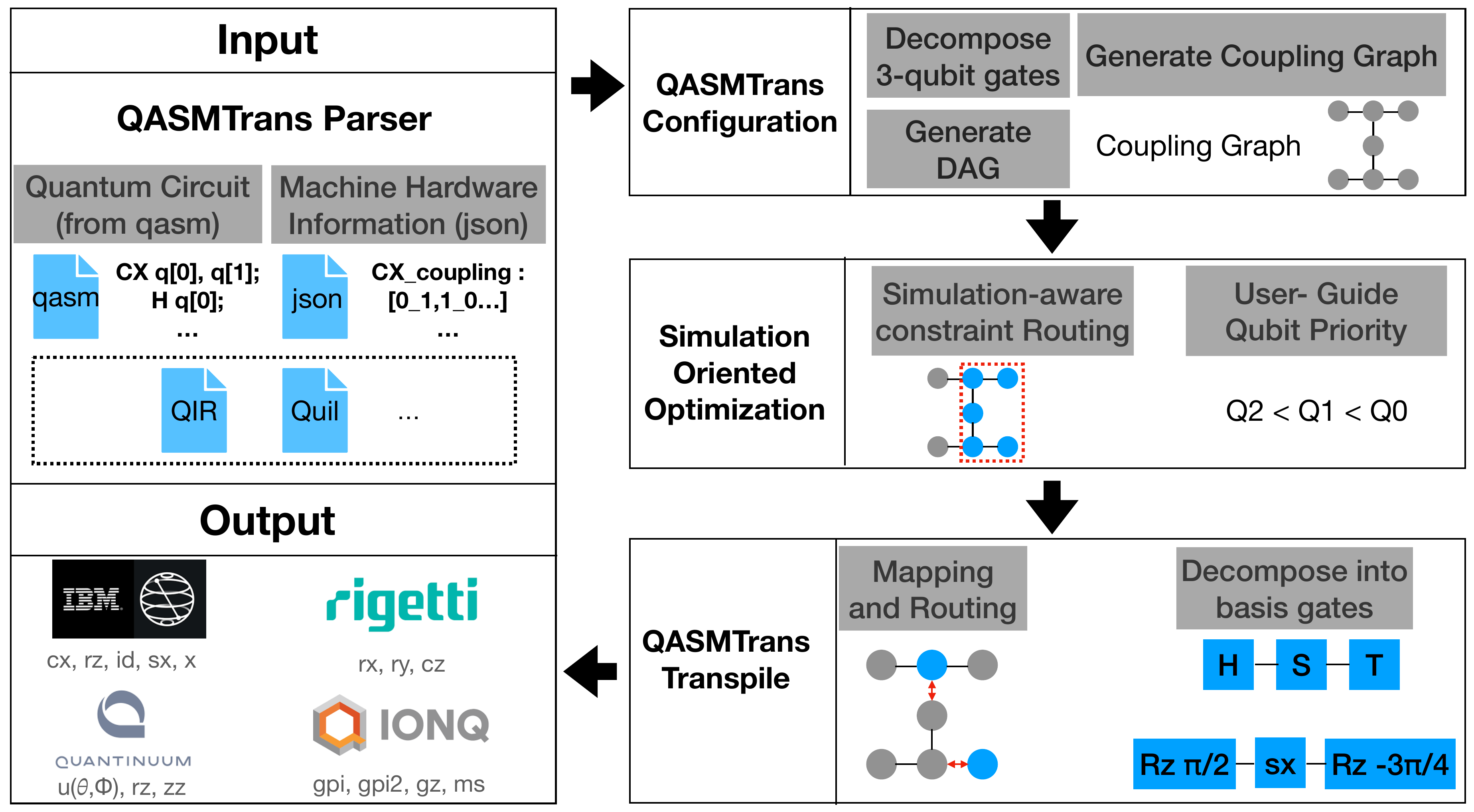}
  \caption{QASMTrans framework, which includes four major components: 1) Input/Output: the Input is the parser that reads in QASM and stores them internally as gate IRs. The Output saves the transpiled circuit in the QASM format. 
  2) Configuration: perform pre-transpilation work such as generating the coupling graph, gate DAG, and 3-qubit gates decomposition. 3) Simulation-oriented Optimization. 4) Transpilation, including mapping, routing, and decomposition into basis gates of the target device.}
  \label{fig:transpiler}
\end{figure}

We elaborate on the QASMTrans transpiler framework in this section, The main structure is shown in Figure~\ref{fig:transpiler}. QASMTrans contains the following main components:
\begin{enumerate}
    \item \textbf{Input/Output (IO)}:
    \begin{itemize}
        \item \emph{Input}: QASMTrans starts with a QASM parser. The parser reads the QASM file, and translates it into a gate IR. Meanwhile, the input module also extracts pertinent hardware details from a JSON file that describes the backend device. We plan to support other input formats such as QIR~\cite{qir} and Quil~\cite{smith2016practical}.
        \item \emph{Output}: Once the transpilation is complete, the circuit is saved to a new QASM file, primed for execution on real quantum hardware.  QIR~\cite{qir} is another format to be supported.
    \end{itemize}    
    \item \textbf{QASMTrans Configuration}:
    \begin{itemize}
        \item \emph{Gate Decomposition}: In this phase, gates with three qubits are methodically broken down into combinations of one- and two-qubit gates. For example, the \texttt{CCX} gate will be decomposed into \texttt{CX} and \texttt{T} gates.
        \item \emph{Directed Acyclic Graph} (DAG): A DAG will be generated for the gates describing the dependency. In the DAG, every vertex represents a physical qubit, whereas each edge represents a coupling link.
        \item \emph{Coupling Graph}: We generate the coupling graph based on the input hardware JSON file, where each vertex represents a physical qubit, and each edge represents the link between qubits. The coupling graph is essential for routing/mapping.
    \end{itemize}
    \item \textbf{QASMTrans Process}:
    \begin{itemize}
        \item \emph{Routing and Mapping}: This involves aligning the given quantum circuit to the specific topology of different quantum machines. To achieve this, we introduce \texttt{SWAP} gates where necessary. As a starting point, we implement the Sabre algorithm~\cite{li+:asplos19} that is also widely used in frameworks such as Qiskit and XACC~\cite{mccaskey2020xacc}.
        \item \emph{Basis Gate Decomposition}: Depending on the desired quantum machines, like Rigetti or Quantinuum, the circuit is further decomposed into the directly executable basis gates of the specific hardware.      
        \item \emph{Pulse Compilation}: After a circuit has been transpiled to the specified machine, a final pass to translate pulses into a pulse schedule is performed.
    \end{itemize}
    \item \textbf{Simulation-Oriented Optimization}:
    \begin{itemize}
        \item \emph{Simulation-Aware Constrained Routing}: To date, many quantum circuits and algorithms are still evaluated in classical simulators. Given the exponential cost of having more qubits to simulate, in QASMTrans, we introduce a method that can limit the number and index of qubits used for the transpilation. This can significantly reduce the transpilation time as well as simulation time.
        \item \emph{Qubit Priority Rescheduling}: Based on user-specified qubit priorities, QASMTrans can optimize and realign the qubit mapping. This is especially useful for distributive classical simulation, as the number of gates over globally shared qubits can be minimized.
    \end{itemize}
\end{enumerate}
\subsection{QASM Parser}

\label{sec:transpiler}
The QASM parser is responsible for parsing the input OpenQASM to the internal gate IR, which will be discussed in more detail below.
\subsubsection{Tokenization using Lexertk}
The parser begins its operation by tokenizing the QASM text, a process that involves breaking down the text into smaller chunks known as tokens. This is achieved by incorporating Lexertk~\cite{lexer}, a high-performance lexer tool written in C++ and distributed through a single C++ header file. The parser of QASMTrans uses Lexertk to scan through the QASM code and break it down into various tokens. Each token is a string of characters that conforms to the \emph{Backus–Naur Form} (BNF), an important notation technique for context-free grammars, defining a set of syntax rules for valid tokens.

\subsubsection{Qubit/Classical Register Management}
The QASM parser automatically flattens the qubit register indices and translates them into a singular range of qubit indices. This process significantly enhances the system's proficiency for transpilation and simulation by replacing the typically used \emph{REG\_NAME[INDEX]} qubit addressing, seen in QASM, with a more streamlined one-dimensional qubit range.
Classical registers are used to store the outcomes of measurements from qubit registers, typically achieved through commands such as:
\[\text{measure }q[0] \rightarrow c[0];\]
In this example, `\emph{q}' denotes a qubit register, and `\emph{c}' denotes a classical register. The QASM parser keeps track of the qubit register remapping, ensuring accurate measurement operations.

\subsubsection{Gate Sets and Abstraction}
\label{sec:gate_ir} 
In the rapidly evolving field of quantum computing, it is crucial to have a robust and flexible system capable of accommodating an extensive range of quantum gates, from the most common to the more advanced. QASMTrans currently supports all the gates (except \texttt{C4X}) defined by the OpenQASM~2 specification, see Table~\ref{tab:gates}.

The parser supports standard gates such as \texttt{Pauli-X}, \texttt{Pauli-Y}, \texttt{Pauli-Z}, \texttt{Hadamard}, \texttt{CNOT}, and \texttt{Toffoli}, as well as parameterized gates like \texttt{RX}, \texttt{RY}, \texttt{RZ}, and \texttt{U} gates. It also accommodates more complex gates like the \texttt{SWAP} gate and the controlled versions of various gates. These are by no means an exhaustive list, and the parser's design allows for easy extension to incorporate additional or newer gate types.

Key to the flexibility and functionality of the QASMTrans is the Gate IR. It is a custom C++ class that encapsulates four crucial aspects of each quantum gate:
\begin{itemize}
    \item \textbf{Gate Name}: Represents the type of quantum gate.
    \item \textbf{Target Qubits}: Specifies the individual qubits upon which the quantum gate operation is performed.
    \item \textbf{Gate Parameters}: Contains the parameters relevant to certain quantum gates.
    \item \textbf{Gate Matrix}: Encapsulates the matrix representation of quantum gate, stored as two arrays --- one for the real and the other for the imaginary components.
\end{itemize}

\subsection{Transpile configuration}
Before the transpilation process, we need to perform some preliminary configuration.
\paragraph{Generation Coupling Graph (full/limited)} Based on the topology of the hardware device, we generate a coupling graph that embeds essential elements such as a distance matrix and an adjacent\_edge\_list. According to the size of the topology, there are two potential approaches: (i) Build the full graph for all the qubits and links. This, however, introduces excessive overhead towards large devices (e.g., the 433-qubit IBM Seattle). (ii) Alternatively, and in most cases, the qubit number of a  circuit is smaller than that of the device. Thus, we can limit the qubits and links of the device (through a partial coupling graph) that are taken into the transpilation consideration, drastically shrinking the search space.

\paragraph{Directed Acyclic Graph (DAG) Generation} From the input circuit, a DAG can be constructed to indicate the gate dependency. For example, nodes with an in-degree of zero can be executed immediately without any dependency. Otherwise, any nodes with non-zero in-degree require all of their parent nodes to be executed beforehand to satisfy the dependency. Considering the efficiency, we only maintain two lists: one is the front list that contains executable gates; the other is the future list comprises gates for future execution.

\paragraph{Decompose three-qubit gates}  In our transpiler, we first decompose all the 3-qubit gates into 1-qubit and 2-qubit gates, given most of the quantum devices use 1-qubit and 2-qubit gates as the basis gate set. For example, the widely used \texttt{Toffoli} gate, or \texttt{CCX} gate, will be decomposed into 6 \texttt{CX} gates and 9 one-qubit gates.

\subsection{Routing and mapping}
After the initial decomposition of 3-qubit gates, the next step is to map the virtual qubits to the physical qubits. Various strategies exist for performing this mapping and routing, with each method optimized for different targets. For instance, Sabre is designed to minimize the number of swaps required \cite{li+:asplos19}. Time-optimal qubit mapping emphasizes minimizing the circuit depth \cite{zhang+:asplos21}. The Noise-Adaptive approach is geared towards minimizing the error of the transpiled circuit \cite{tannu+:asplos19}. 

In QASMTrans, we use Sabre as the primary approach, due to its significant advantages in compilation time compared to the others. The major remaining overhead in Sabre routing and mapping includes:
1) After the execution of each gate, we need to update the DAG and regenerate the new front list of gates with in-degree equals to 0 in the DAG (if the gate is in the execution list, its dependency must have already been satisfied and it is ready for execution). The original Sabre method traverses the entire circuit (i.e., all DAG nodes) and identifies the gates that are ready to be executed. As QASMTrans is designed to address very deep circuits, this cost of traversing can be significant. To accelerate this process, we propose to keep the same front layer for each step, but only delete the nodes that are just executed, and fetch any new gates whose dependencies are just resolved through the step. Given that in each time step, only $n$ gates can be simultaneously executed, our proposed optimization can essentially reduce the searching cost of Sabre from $O(G)$ where $G$ is the total number of gates, to $O(n)$ where $n$ is the number of qubits. As the circuit gate number scales this speed up becomes more pronounced.

2) When a \texttt{SWAP} operation is required, selecting the appropriate \texttt{SWAP} requires the calculation of all possible swaps, creating a large search space and significant overhead. This is particularly the case for large machine targets. Consequently, we propose a new method that prunes the pool of \texttt{SWAP} candidates by constraining the physical qubit area. This will be discussed in Section~\ref{subsec:simu-oriented_opt}.

\subsection{Decompose to basis gates}
Here we perform the final decomposition towards the basis gates of the device after routing and mapping. The main consideration is efficiency and simplicity, as decomposing into basis gates before routing and mapping can drastically enlarge the search space during routing and mapping. 

The decomposition here is a translation from general gates to the targeted basis gates. The basis gate set for IBMQ, Rigetti, Quantinuum, and IonQ can be found in Table~\ref{tab:basis_gates}. The detailed translation rules can be found in the open-source code of QASMTrans.

\subsection{Statistics}
Based on the circuits, QASMTrans can print out the following circuit metrics based on statistics of the quantum gates in the circuit. The detailed definition can be found in \cite{li2023qasmbench}.
\begin{itemize}
\item \textbf{Circuit Depth} represents the minimum count of time-evolution steps needed to complete a quantum circuit, calculated based on standard QASM gates.
\item \textbf{Gate Density} indicates the utilization of gate slots during the time evolution of a quantum circuit, similar to pipeline occupancy in classical processors.

\item \textbf{Retention Lifespan} quantifies the maximum longevity of a qubit within a system. Its relationship with the T1 and T2 time of the device dictates the feasibility of the circuit execution on the targeted device.
\item \textbf{Measurement Density} evaluates the importance of measurement operations in a circuit, with respect to the overall induction fidelity.

\item \textbf{Entanglement Variance} measures the balance of entanglement across the qubits for a circuit. It indicates the level of connectivity and the potential error reduction through an advanced transpiler.

\end{itemize}

\subsection{Simulation-oriented Optimization} \label{subsec:simu-oriented_opt}

As mentioned, most of the contemporary circuit inductions are still performed through classical simulations. In QASMTrans, we propose two classical simulation-oriented optimizations during transpilation to generate circuits that can be simulated more efficiently. 

\paragraph{Constrained qubit routing/mapping} During the routing and mapping phase, instead of considering all the physical qubits of the device, we limit the number and coupling of qubits that will be considered during the transpilation, based on the number of virtual qubits used in the circuit. This is achieved by first adopting the isomorphic algorithm to find the most relevant connected graph from the hardware architecture, using the number of virtual qubits as input. The qubits of the obtained graph should contain equal or more qubits than the circuit virtual qubits, but less or equal to the number of physical qubits in the device. We then refer to the routing algorithm as normal. Although constrained routing and mapping with partial graphs can lead to more swaps, the benefit of simulating fewer qubits can extraordinarily speed up the transpilation process.

\paragraph{User-guided qubit prioritization}
Another simulation-oriented optimization is to enforce user-defined qubit prioritization. Users can specify a priority order such as $q3<q1<q0<q2$, then for classical simulation, we can perform a qubit remapping with respect to this partial order. This is achieved by counting the number of gates performed on each qubit, sorting, and then re-indexing the qubits to assign high-priority qubits to perform more gates. For example, if $q2$ shows the best performance or least error rate, which is set to have the highest priority, the qubit with the most number of gates can be remapped to it. On the other hand, if the coefficients of $q3$ are distributed across multiple nodes for large-scale distributive simulation (i.e., a global qubit), because of the overwhelming cost from inter-node communication, it is set to the lowest priority, we would want the least number of gates to be mapped to $q3$.

%% file: Pulses.tex
\section{Pulse Compilation} \label{sec:pulse}
\begin{figure*}[h!]
    \centering
    \begin{minipage}[c]{0.32\textwidth}
        \centering
        \includegraphics[width=\linewidth]{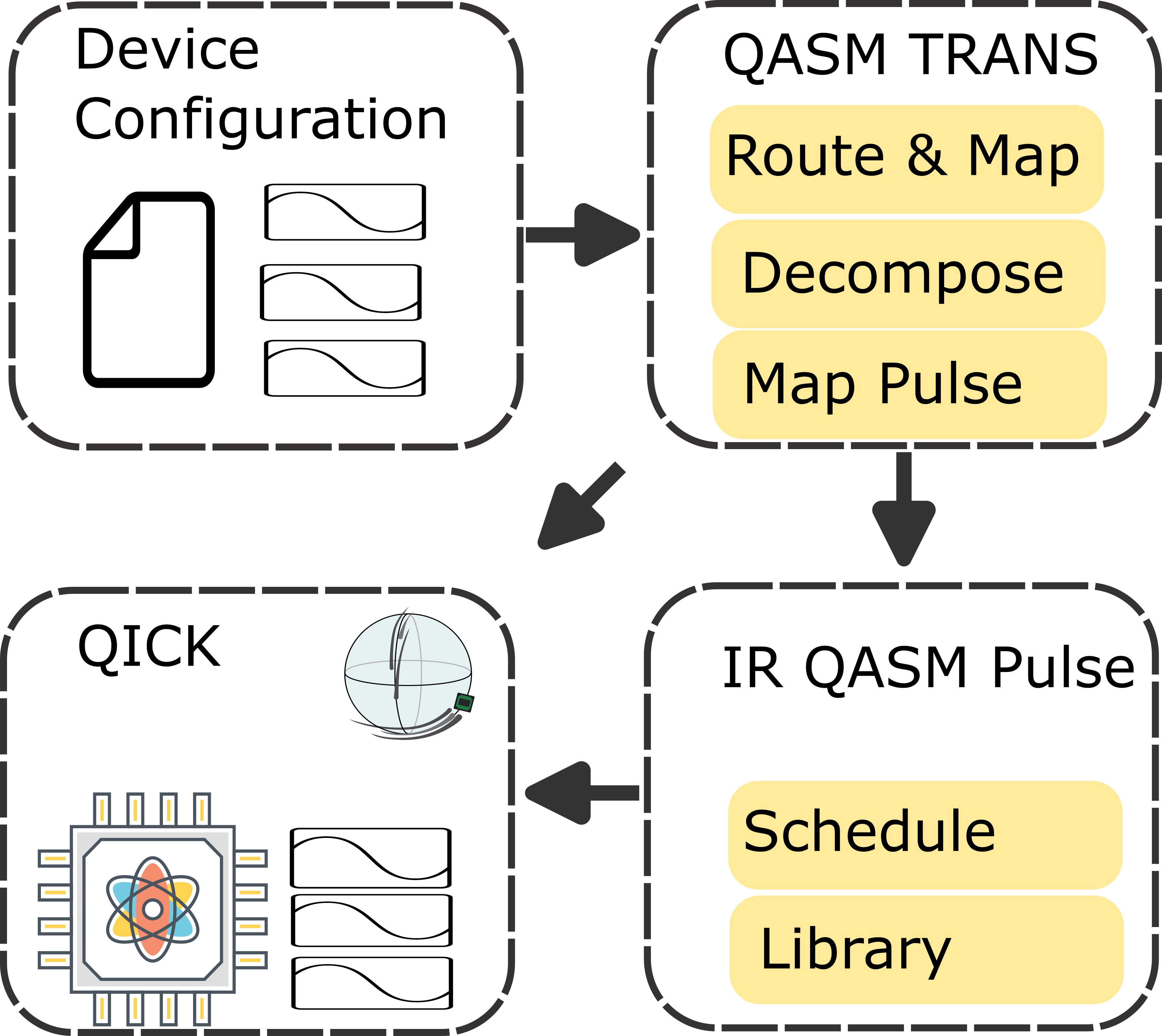}
    \end{minipage}\hfill
    \begin{minipage}[c]{0.32\textwidth}
        \centering
        \includegraphics[width=\linewidth]{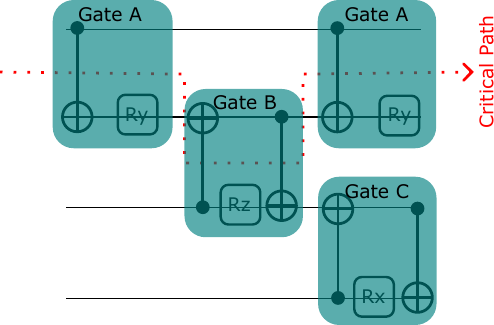}
    \end{minipage}\hfill
    \begin{minipage}[c]{0.32\textwidth}
        \centering
        \includegraphics[width=\linewidth]{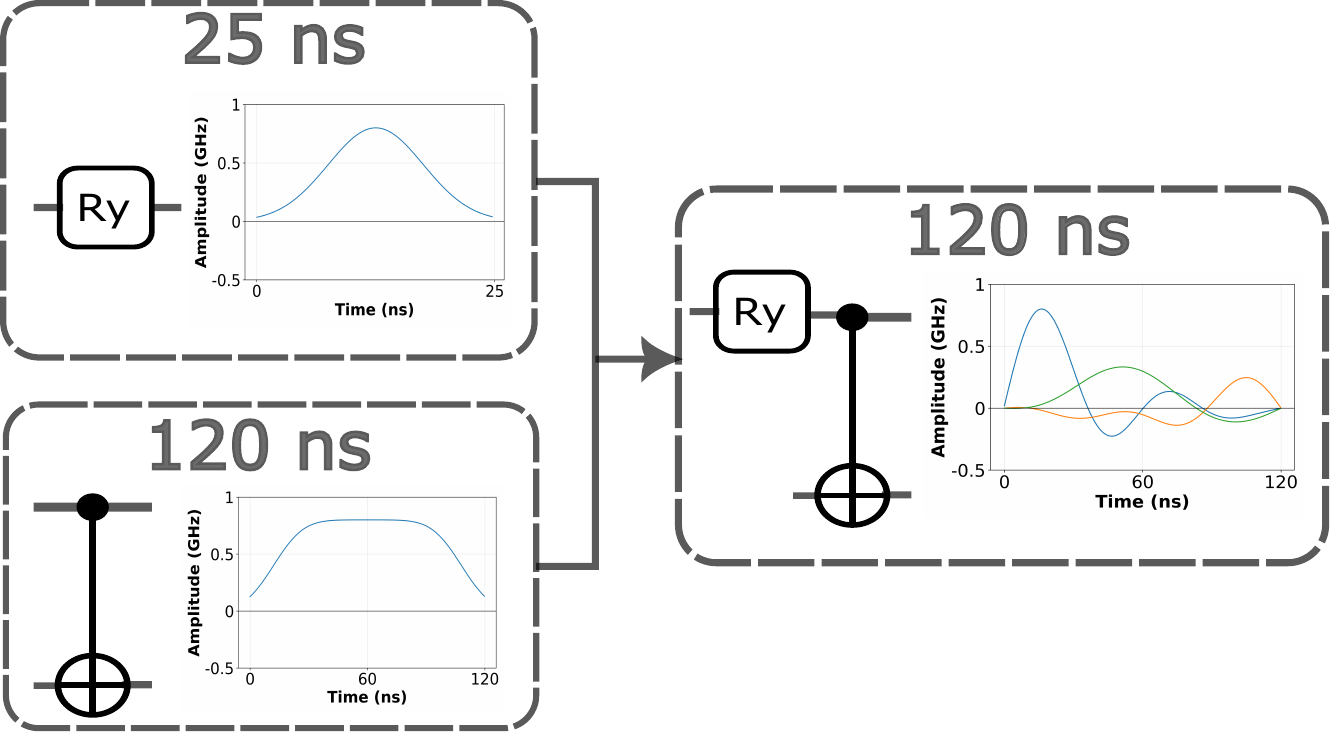}
    \end{minipage}
    \\[2pt]
    \makebox[0.32\textwidth][c]{\textbf{a)}}\hfill
    \makebox[0.32\textwidth][c]{\textbf{b)}}\hfill
    \makebox[0.32\textwidth][c]{\textbf{c)}}
    \caption{QASMTrans pulse compilation pipeline with an Application-tailored Gate Set scheme: a) QASMTrans program flow of compilation to a device pulse schedule, b) example of logical level gates added to the AGS along the critical path, c) optimization of new gate for AGS.}
    \label{fig:merge_overview}
\end{figure*}

While the core QASMTrans transpilation framework (Section \ref{sec:transpiler}) transforms logical quantum circuits into device-compatible gate sequences, achieving optimal performance on near-term quantum hardware requires extending compilation down to the pulse level. Pulse-level control enables fine-grained optimization of quantum operations but remains underserved by existing transpilation frameworks. The Amazon Braket SDK supports QASM 3.0 pulse compilation only for Rigetti pulses, Qiskit restricts pulse operations to IBM hardware, and XACC only supports compilation of quantum programs to its built-in pulse simulator\cite{aws_braket_pulse_control,alexander+:qiskitpulse20,9347736}.

QASMTrans addresses these limitations through an end-to-end compilation pipeline from device configurations and QASM circuits to calibrated pulse schedules. This pulse-level compilation layer enables three critical advances: (1) direct hardware integration through QICK for embedded quantum systems, eliminating the overhead of external pulse generation tools, (2) portable pulse file generation validated through an integrated QuTiP-based simulator, and (3) an application-tailored gate set (AGS) that reduces circuit latency by synthesizing custom pulse sequences for high-impact pre-transpiled gates along the critical path. These capabilities complement the gate-level transpilation passes while maintaining the framework's emphasis on compilation speed and deployment flexibility, enabling rapid calibration cycles and adaptive algorithm execution on a quantum device.
\subsection{QICK Integration}

Among available control and calibration stacks (e.g., Q-CTRL's Boulder Opal and LBNL's QuBiC), we adopt the Quantum Instrumentation Control Kit (QICK) to maintain an RFSoC-native, open, and fully modifiable control/readout path that minimizes external instrumentation while supporting low-latency adaptive execution via QASMTrans \cite{Stefanazzi_2022,xu_qubic_2021}. Designed for FPGA and embedded deployments, QICK utilizes high-speed DACs and ADCs with numerically controlled oscillators, mixers, and real-time pulse sequencers, all accessible through a Python API. The minimal transpilation overhead of QASMTrans enables adaptive quantum circuits with embedded, real-time compilation. Future work will explore integration with LBNL's QuBiC to leverage its calibration and benchmarking workflows.

QASMTrans supports two emission pathways to QICK hardware, illustrated in Figure~\ref{fig:merge_overview}(a). In the first mode, QASMTrans generates JSON pulse schedule files containing calibrated pulse schedules, which are subsequently parsed and executed via an included Python script. In the second mode, users invoke the \verb|--emit| flag through the command-line interface, triggering direct pulse emission to the QPU through QASMTrans's \verb|pybind|-based API integration with QICK \cite{Stefanazzi_2022}. This dual-mode architecture balances flexibility for offline analysis with low-latency requirements for real-time control.

\begin{figure}[h]
  \centering
  \includegraphics[width=1\linewidth]{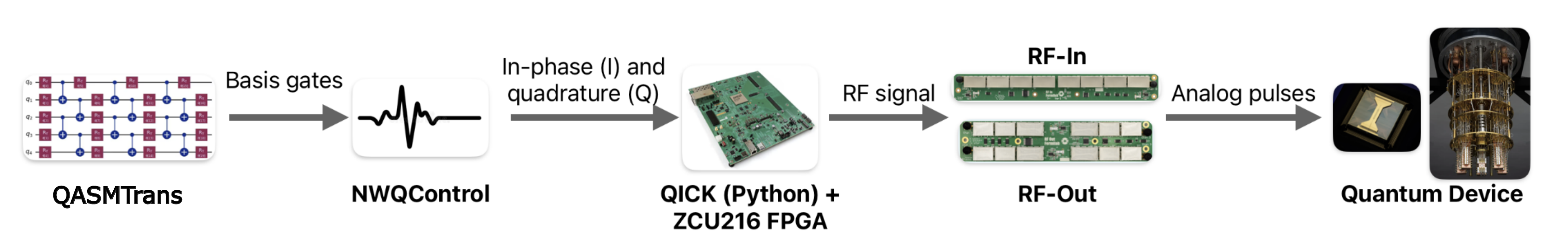}
  \caption{QASMTrans is directly compatible with NWQ-Control. QASMTrans produces pulse schedules and QASM2 files that will integrate directly into the stack allowing the control system to calibrate the testbed and execute quantum gates enabling closed-loop optimization.}
  \label{fig:nwq-control}
\end{figure}

QASMTrans is currently designed to interface directly with QICK, however, to realize the AGS scheme we will design our own control stack to automate the calibration of the tailored gates described in NWQWorkflow \cite{li2026nwqworkflownorthwestquantumworkflow}. NWQ-Control will fill this gap, with built-in tools for qubit calibration and pulse generation. The future end-to-end workflow is shown in Fig. \ref{fig:nwq-control}, QASMTrans outputs QASM files and pulse schedules which are fed into NWQControl which handles the calibration and execution of the quantum programs at the pulse level using the QICK controller to a real quantum device. 

\subsection{Quantum Pulse Simulator with QuTiP}

To validate pulse schedules generated by QASMTrans, we implement a pulse-level simulator using Qutip \cite{Johansson_2012}. The simulator operates on the same device configuration files used for pulse generation, ensuring consistency between compilation and validation. We demonstrate the end-to-end workflow using calibration data from a Rigetti QPU accessed via AWS Braket, modeling a simplified 9-qubit system based on Rigetti's transmon architecture \cite{PhysRevApplied.16.024050}.

The Rigetti Ankaa 3 QPU employs flux-tunable qubits with tunable couplers to suppress crosstalk and minimize leakage through two-qubit interactions. The system Hamiltonian is expressed as:

\begin{equation}
\begin{aligned}
H(t)=&\sum_{i=1}^{N}\Big(\tfrac{1}{2} I_i(t)\,\sigma_x^{(i)}+\tfrac{1}{2} Q_i(t)\,\sigma_y^{(i)}\Big)
&+\sum_{\langle i,j\rangle}\tfrac{1}{2} J_{ij}(t)\big(\sigma_x^{(i)}\sigma_x^{(j)}+\sigma_y^{(i)}\sigma_y^{(j)}\big),
\end{aligned}
\label{eq:H}
\end{equation}

\noindent where $\langle i,j\rangle$ denotes coupled qubit pairs in the device topology, $\sigma_x$ and $\sigma_y$ are Pauli operators, $I_i(t)$ and $Q_i(t)$ represent I/Q modulated drive waveforms for qubit $i$, and $J_{ij}(t)$ is the flux drive mediating two-qubit interactions.

Rigetti's native gate set comprises quantized single-qubit rotations $R_x(-\pi,-\pi/2, \pi/2, \pi)$ implemented via I/Q control, virtual $R_z(\theta)$ gates realized through phase tracking, and the $\text{iswap}$ entangling gate \cite{PhysRevApplied.16.024050}. To minimize calibration overhead, the basis set quantizes Clifford operations while maintaining parameterized virtual-$Z$ gates for non-Clifford rotations. We implement virtual-$Z$ gates through frame tracking, maintaining a phase map that adjusts the reference phase for each subsequent single-qubit $R_x$ pulse.

The $\text{iswap}$ gate arises from the $(XX+YY)$ coupling term in Equation~\ref{eq:H}. Rigetti's calibration employs two synchronized flux pulses: one tunes a qubit's frequency to enable resonant interaction, while the other modulates the coupler flux to control interaction strength \cite{PhysRevApplied.16.024050}. For computational efficiency, our simulator neglects qubit frequency detuning, modeling only the effective coupling dynamics.

Decoherence is incorporated via the Lindblad master equation, utilizing collapse operators $\{\sigma_-,\sigma_+, \sigma_z\}$ with decay rates $\kappa_i$ (relaxation) and $\gamma_i$ (dephasing):

\begin{equation}
\begin{aligned}
\dot{\rho}(t)=&-i\,[H(t),\rho(t)]
&+\sum_{i=1}^{N}\kappa_i\Big(\sigma_-^{(i)}\rho(t)\sigma_+^{(i)}-\tfrac{1}{2}\{\sigma_+^{(i)}\sigma_-^{(i)},\rho(t)\}\Big)
&+\sum_{i=1}^{N}\tfrac{\gamma_{i}}{2}\Big(\sigma_z^{(i)}\rho(t)\sigma_z^{(i)}-\rho(t)\Big).
\end{aligned}
\label{eq:lindblad}
\end{equation}

The simulator accepts user-provided device files specifying topology, $T_1/T_2$ coherence times, and single- and two-qubit Hamiltonians, ensuring extensibility to alternative architectures. We employ the Rigetti platform for demonstration due to the accessibility of calibration data through AWS Braket. In subsequent sections, we apply this simulator to validate the AGS scheme and benchmark algorithm performance.

\subsection{Synthesis and Simulation of the Application-tailored Gate Set (AGS)}

Limited qubit coherence times constitute a primary obstacle to practical NISQ computation \cite{clarke2008superconducting, rigetti2012superconducting}. One mitigation strategy involves identifying frequently occurring pre-transpiled gates and consolidating them into single calibrated pulse with reduced latency, we denote this the Application-tailored Gate Set or AGS. QASMTrans implements this capability through latency-aware pre-transpiled gate ranking: pre-transpiled gates are ranked by their aggregate contribution to the circuit's critical path, enabling users to trade calibration time for reduced execution latency. Figure~\ref{fig:merge_overview} illustrates the AGS workflow. In this work, all results are obtained with the Qutip-based pulse simulator using device configuration files derived from Rigetti calibration data.

AGS candidates are selected using critical path information computed during the routing and mapping pass. We rank one- and two-qubit pre-transpiled gates by their cumulative latency contribution along the critical path, enabling candidate identification in $O(N)$ time without degrading overall compilation performance. To maintain practical calibration overhead, we restrict AGS to one- and two-qubit gates. After QASMTrans identifies high-impact gates, we optimize according to the selected ansatz. We use the L-BFGS-B algorithm from SciPy to optimize the process fidelity produced from evolving the pulse through our QPU simulation \cite{virtanen2020scipy}. This can be easily computed from QuTip by computing the propagator and average fidelity, we show the relevant expressions in Eq. \ref{eq:unitary_propagator} and Eq.  \ref{eq:average_gate_fidelity_unitary} where $\mathcal{T}$ is the time-ordering operator which is necessary for time varying controls \cite{lambert2026qutip}.

\begin{equation}
\begin{aligned}
U(T)
=&\; \mathcal{T}\exp\!\left(-i\int_{0}^{T} H(t)\,dt\right),
\end{aligned}
\label{eq:unitary_propagator}
\end{equation}

\begin{equation}
\begin{aligned}
F_{\mathrm{avg}}(U,U_{\mathrm{tgt}})
=&\; \frac{\left|\mathrm{Tr}\!\left(U_{\mathrm{tgt}}^{\dagger}U\right)\right|^{2}+d}{d(d+1)}
\end{aligned}
\label{eq:average_gate_fidelity_unitary}
\end{equation}

To assess the experimental feasibility of the proposed approach, we restrict the pulse ansatz to schemes that have been experimentally demonstrated for high-fidelity single- and two-qubit operations. For single-qubit control, we consider two AGS ansatz, depending on whether the hardware platform supports virtual \(Z\) gates. When virtual \(Z\) gates are available, arbitrary single-qubit unitaries can be implemented using an Euler-angle decomposition, requiring only a single physical pulse together with frame updates that realize the virtual \(Z\) rotations. When virtual \(Z\) gates are not supported, arbitrary single-qubit unitaries can instead be robustly implemented using two phase-shifted physical pulses. The latter construction is particularly important when combined with the AshN scheme for two-qubit gates, since the implementation of generic two-qubit unitaries precludes the use of virtual \(Z\) gates due to noncommutativity with nonlocal interactions \cite{PhysRevA.96.022330}. Accordingly, throughout this work we primarily adopt the second decomposition in Eq.~\ref{eq:su2_decompositions} to ensure compatibility with the AshN-based realization of two-qubit gates.

\begin{equation}
\begin{aligned}
U
=&\; R_z(\alpha)\,R_x(\beta)\,R_z(\gamma) \\
=&\; R_{\phi_2}(\theta_2)\,R_{\phi_1}(\theta_1),
\end{aligned}
\label{eq:su2_decompositions}
\end{equation}

The implementation of arbitrary two-qubit gates in a robust and hardware-efficient manner is considerably more challenging. To address this, we employ the AshN pulse scheme, which has been theoretically developed and experimentally validated on superconducting quantum processors \cite{10.1145/3620665.3640386, chen2025efficient}. This scheme provides an analytic mapping from an arbitrary two-qubit unitary to a single entangling pulse, up to local single-qubit corrections, which can be efficiently realized using the single-qubit pulse constructions described above. The AshN scheme is compatible with any hardware platform capable of realizing an iSWAP-like interaction \cite{chen2025efficient}. Its advantages are twofold: first, the existence of an analytic pulse solution yields high-quality initial parameters, substantially reducing calibration overhead; second, as experimentally demonstrated in Ref.~\cite{chen2025efficient}, AshN pulses can realize two-qubit gates with shorter duration than decompositions into a processor’s native gate set, thereby reducing circuit latency and improving overall execution fidelity.

\begin{figure*}[h!] \centering \begin{minipage}[t]{0.5\textwidth} \centering \includegraphics[width=\linewidth]{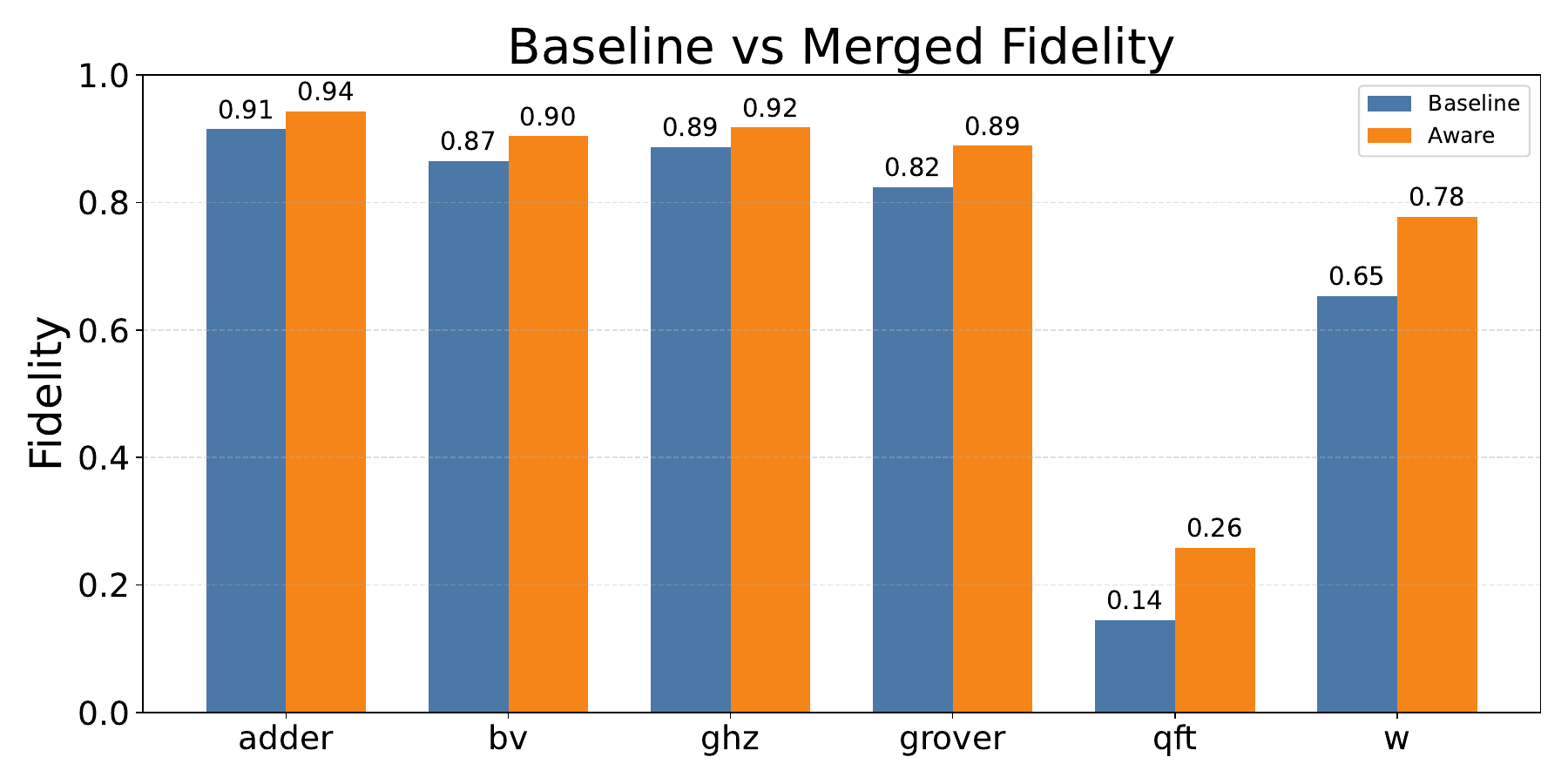}\\[2pt] \textbf{a)} \end{minipage}\hfill \begin{minipage}[t]{0.5\textwidth} \centering \includegraphics[width=\linewidth]{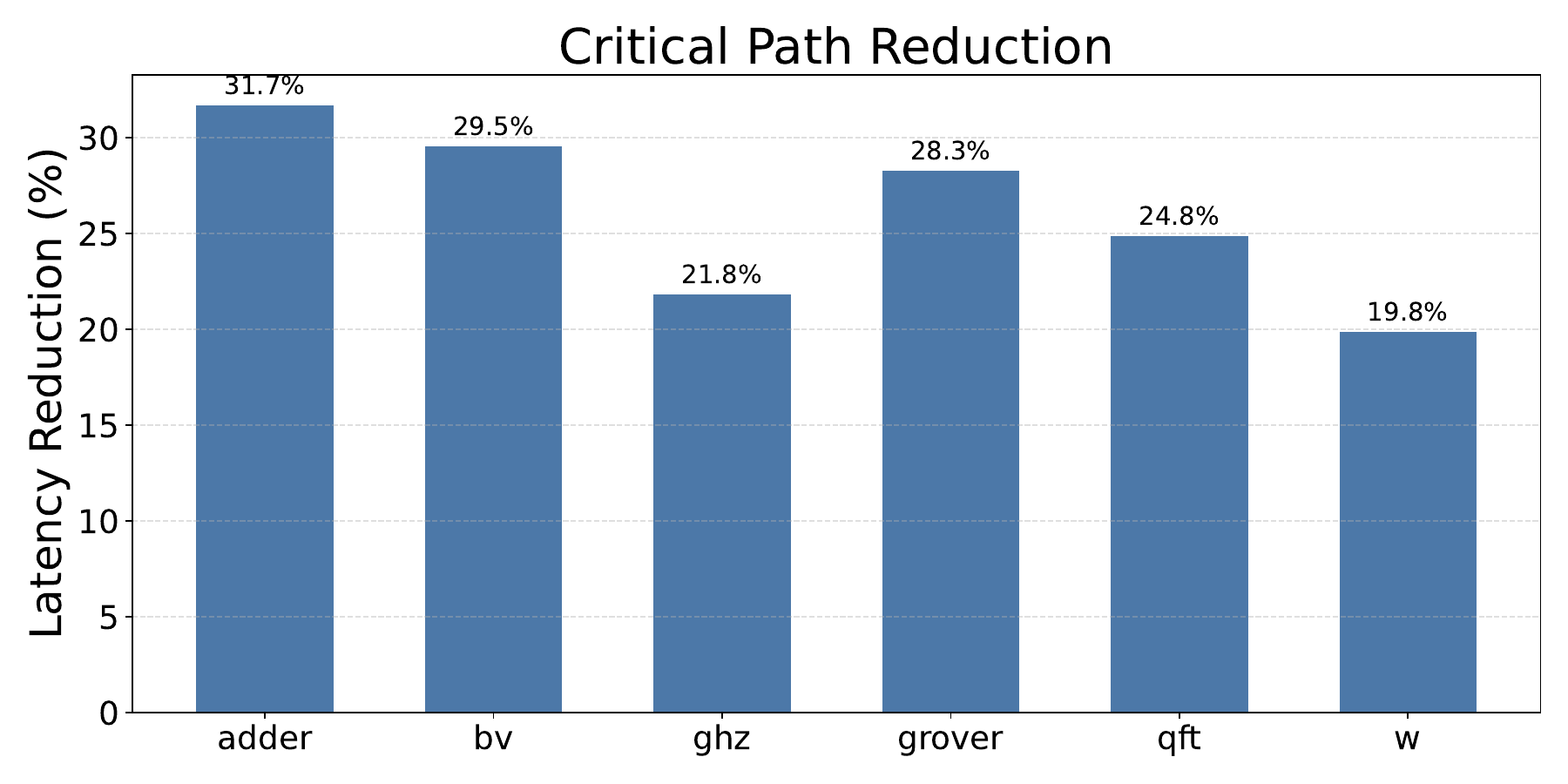}\\[2pt] \textbf{b)} \end{minipage}\hfill \caption{Demonstration of calibrating a native CNOT gate with the AshN scheme. We simulate a 7 qubit Rigetti like device with native rx and iswap gates. In the baseline method we compile to rx and iswap pulses, in the application aware we include a calibrated AshN CNOT gate which decreases the circuit latency, thus increasing circuit execution quality. a) Consitently the application aware compilation increases the simulated fidelity of the executed circuit. b) Reduction of critical path due to the introduction of the single ashn cnot gate. We assume a rx gate of 10 ns and a iswap gate time of 40 ns, the calibrated ashn CNOT gate has a total time of 100 ns, however this is still low enough to beat the decomposition of the CNOT gate to rx and iswap pulses. Each native gate is calibrated to near perfect precision, so circuit latency is the main factor of circuit infidelity.} \label{fig:merge_results} \end{figure*}

The AshN scheme is based on the KAK decomposition, which states that any two-qubit unitary \(U \in SU(4)\) can be expressed as \(U = \lambda (K_1 \otimes K_2) U_w (K_3 \otimes K_4)\), where \(\lambda \in \{1,i\}\), \(K_j \in SU(2)\) are local single-qubit unitaries, and \(U_w(a,b,c) = \exp[i(a\,XX + b\,YY + c\,ZZ)]\) corresponds to a unique point \((a,b,c)\) in the Weyl chamber. The Weyl chamber coordinates fully characterize the nonlocal content of the two-qubit gate, while the local unitaries account for equivalence under single-qubit operations. For an arbitrary two-qubit unitary, the AshN scheme provides an analytic procedure to synthesize a pulse that realizes the target Weyl chamber coordinates. A key assumption of the AshN scheme is that the entangling interaction strength remains constant throughout the pulse duration. Consequently, there exists a minimum interaction time required to reach a given point in the Weyl chamber \cite{10.1145/3620665.3640386}. The resulting pulse consists of a parallel set of control fields, parameterized by the amplitudes of the \(X\)-type drives applied to each qubit and a shared detuning term, as shown in Eq.~\ref{eq:H_twoqubit}. In total, three continuous parameters specify the gate, with the total interaction time optionally treated as a fourth degree of freedom.

\begin{equation}
\begin{aligned}
H
=&\; \frac{\Delta}{2}\left(ZI + IZ\right)
    + \frac{g}{2}\left(XX + YY\right)
    + \frac{\Omega_1}{2}\,XI
    + \frac{\Omega_2}{2}\,IX,
\end{aligned}
\label{eq:H_twoqubit}
\end{equation}

To validate AGS scheme we utilize the QPU simulator discussed in the previous section to simulate a 7 qubit subset of a Rigetti QPU. The Rigetti QPU naturally supports the iswap interaction and will work well with the AshN scheme. For the baseline we calibrate the Rx and iswap pulses to near perfect fidelity, as to isolate the circuit fidelity drop to qubit coherence time rather than uncalibrated gates. After running the baseline transpiled circuit we compile a single CNOT gate from the AshN scheme recommended from the QASMTrans gate candidate list, then retranspile in qasmtrans with the new pulse schedule. We find that for all circuits tested this decreased the critical path and subsequently increased simulated circuit fidelity. Circuit fidelity is calculated from the ideal circuit simulator in NWQ-Sim vs the output density matrix state from QuTiP with the following fidelity relation $F=\operatorname{Tr}\!\left(\rho\, |\psi\rangle\langle\psi| \right)$. Experimental results are summarized in Fig. \ref{fig:merge_results}.

 When implemented on real hardware, the application specific gates would be part of a online pre-calibration step. The ansatz used for the AGS scheme has already been realized previously and will be feasible on real systems with pre-calibration. We outline some established calibration pipelines: DRAG for leakage-suppressed single-qubit controls \cite{Motzoi2009DRAG}; RB/ORBIT as a closed-loop objective for tuning amplitudes, detunings, and timings \cite{Kelly2014ORBIT} simultaneous RB to diagnose addressability and crosstalk before promoting a AGS pulse \cite{Gambetta2012SRB}; RESTLESS and high-throughput acquisition to accelerate convergence \cite{Rol2017Restless,Werninghaus2021HighSpeed}; and gate-specific routines such as cross-resonance cancellation and echoing for robust two-qubit operation \cite{Sheldon2016CRtuneup}. Building off of these techniques, a QPU control stack can calibrate QASMTrans suggested gates to realize the AGS scheme.

%% file: Adaptive.tex
\section{ADAPTIVE TRANSPILATION} \label{sec:adaptive}

\begin{figure*}[t]
    \centering
    \begin{minipage}[c]{0.49\textwidth}
        \centering
        \includegraphics[width=0.7\linewidth]{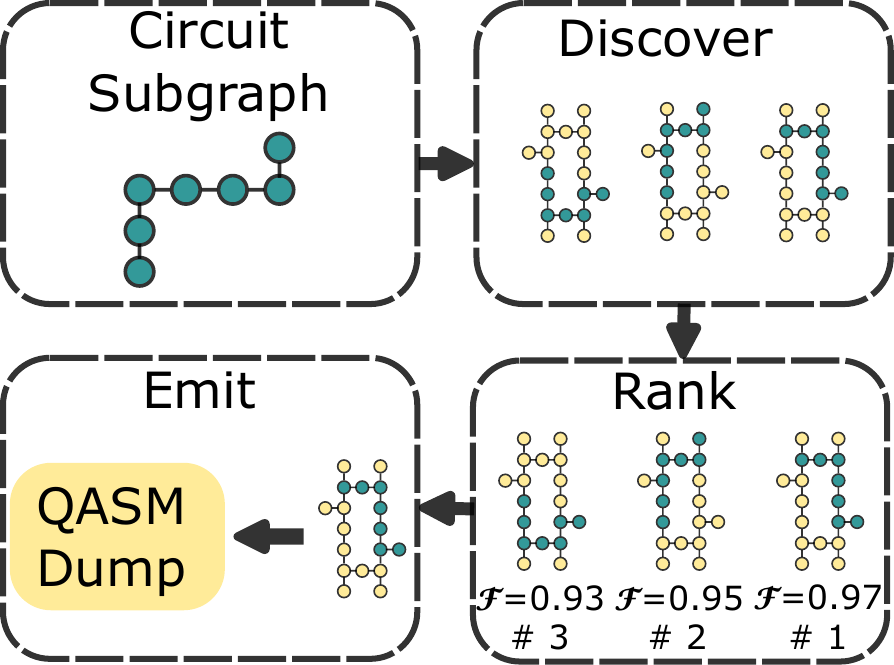}
    \end{minipage}\hfill
    \begin{minipage}[c]{0.49\textwidth}
        \centering
        \includegraphics[width=0.8\linewidth]{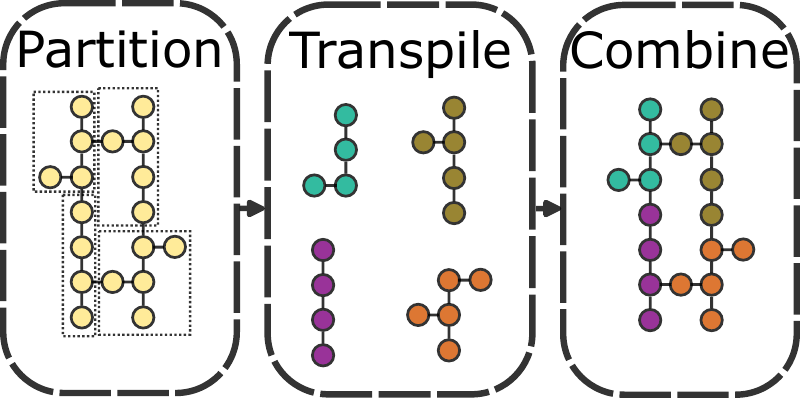}
    \end{minipage}
    \\[2pt]
    \makebox[0.49\textwidth][c]{\textbf{a)}}\hfill
    \makebox[0.49\textwidth][c]{\textbf{b)}}
    \caption{(a) The Noise-Adaptive Pass runs after routing and mapping, using the fixed coupling graph to perform an isomorphic search with VF2++. It enumerates feasible chip layouts, ranks them by a user-defined or critical-path heuristic, selects the optimum, and forwards it to the decomposition pass. (b) Space sharing with QASMTrans.}
    \label{fig:noise_adapt_sp∂acesharing}
\end{figure*}

Building upon the foundational transpilation pipeline (Section \ref{sec:transpiler}) and pulse-level optimization (Section \ref{sec:pulse}), Section \ref{sec:adaptive} introduces adaptive transpilation strategies that leverage device calibration data for both single-circuit placement and concurrent multi-circuit execution. While earlier sections focused on transforming individual circuits efficiently, the techniques presented here address resource allocation and noise mitigation across the full device topology, demonstrating how QASMTrans's fast compilation enables sophisticated device-aware optimization without sacrificing the real-time performance required for adaptive quantum algorithms.

Many quantum transpilers have implemented adaptive transpilation strategies. Notably, Qiskit introduced Mapomatic, which uses subgraph isomorphism to identify alternative circuit placements on hardware topologies \cite{PRXQuantum.4.010327}. We adopt a Mapomatic-inspired approach but enhance it with heuristics that prioritize critical-path optimization rather than global circuit fidelity, enabling faster transpilation with targeted error mitigation.

Beyond single-circuit placement, QASMTrans introduces \emph{space sharing}: a partitioning pass that decomposes the hardware coupling graph into isolated subchips and maps multiple circuits concurrently (Figure~\ref{fig:noise_adapt_sp∂acesharing}(b)). Space sharing and adaptive placement are coupled as follows. First, the device is partitioned using calibration-aware metrics to reduce crosstalk and concentrate high-quality edges within each region. Second, each region is treated as an independent target for the  Mapomatic-style search, where our latency-aware heuristic ranks candidate embeddings by the average error along the circuit's critical path. Finally, the compiled circuits are stitched together. This integrated design separates concerns—partitioning, placement, and scheduling—while allowing each stage to reuse calibration data and critical-path information to minimize compilation time and expected execution latency.

\subsection{Mapping Pass}

The adaptive transpilation pass operates in two stages. First, we construct the interaction graph representing the qubit connectivity of the routed circuit. Second, we apply the VF2++ subgraph isomorphism algorithm to identify all valid mappings of this interaction graph onto the target hardware topology \cite{juttner2018vf2++}. The VF2++ algorithm is implemented using the LEMON C++ graph library \cite{Dezso2011LEMON}.

Figure~\ref{fig:noise_adapt_sp∂acesharing}(a) illustrates the noise-adaptive placement procedure. After enumerating candidate isomorphic subgraphs, we rank each mapping according to the average gate error along the circuit's critical path—defined as the longest dependency chain in terms of gate latency. For a given mapping, we evaluate the heuristic:
\begin{equation}\label{eq:cp-avg-error}
\bar{E}_{\mathrm{cp}} = \frac{1}{L_{\mathrm{cp}}}\sum_{g \in P_{\mathrm{cp}}} \mathcal{E}_{g},
\end{equation}
where $L_{\mathrm{cp}}$ is the critical path length, $P_{\mathrm{cp}} = \{g_{1},\dots,g_{L_{\mathrm{cp}}}\}$ denotes the gates comprising the critical path, and $\mathcal{E}_{g}$ is the error rate of gate $g$ under the proposed mapping. The mapping that minimizes $\bar{E}_{\mathrm{cp}}$ is selected for execution. 

\begin{figure}[h!]
    \centering
    \begin{minipage}[c]{0.49\linewidth}
        \centering
        \includegraphics[width=\linewidth]{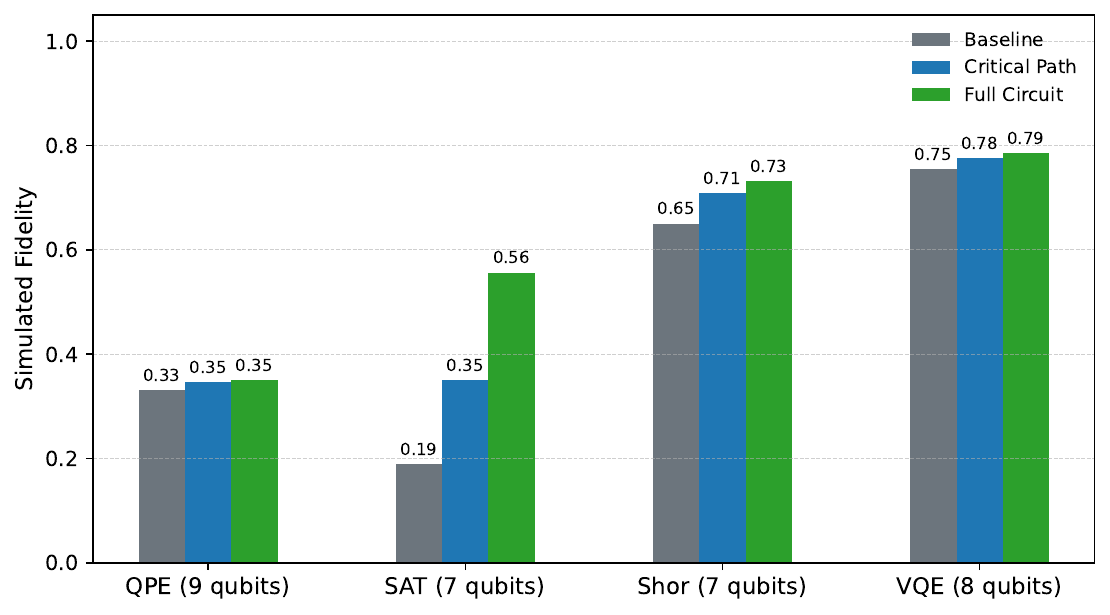}
    \end{minipage}\hfill
    \begin{minipage}[c]{0.49\linewidth}
        \centering
        \includegraphics[width=\linewidth]{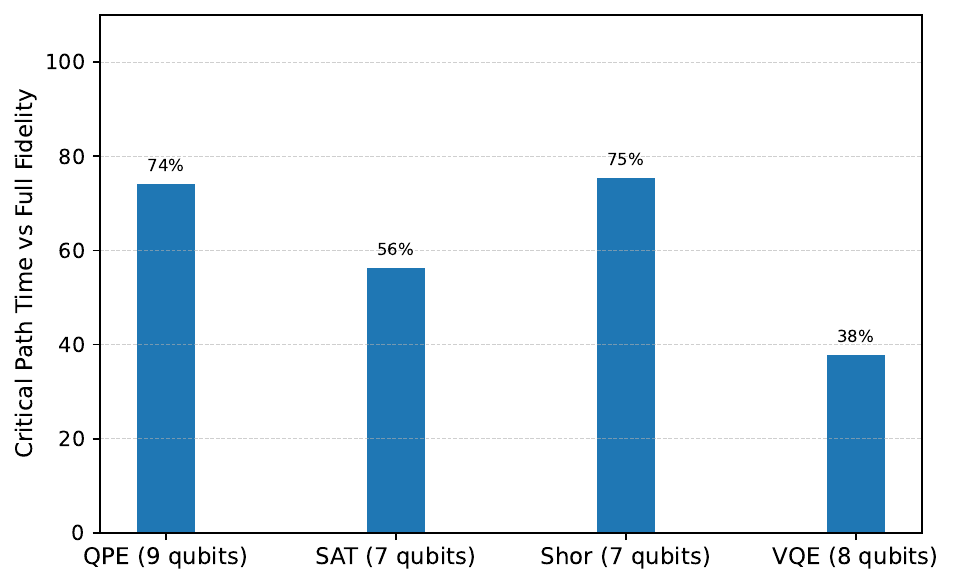}
    \end{minipage}
    \\[2pt]
    \makebox[0.49\linewidth][c]{\textbf{a)}}\hfill
    \makebox[0.49\linewidth][c]{\textbf{b)}}
    \caption{Critical Path Heuristic comparison to the default mapomatic heuristic. a) We observe similar fidelity outcomes for heuristic computed from the critical path over the whole circuit. b) We observe a decreased Mapomatic pass time with this heuristic demonstrating increased efficiency.}
    \label{fig:mapomatic_implementation}
\end{figure}

In Figure~\ref{fig:mapomatic_implementation} we test the noise-adaptive transpile pass and find consistently improved results using the NWQ-Sim quantum circuit simulator  \cite{li2020density}. For most circuits the critical path heuristic can reduce circuit transpilation time with similar performance improvement in the simulated QPU.

\subsection{Space Sharing}

Given a hardware coupling graph with calibration data, QASMTrans partitions the device so multiple circuits can execute concurrently, each confined to its own subchip. Figure~\ref{fig:noise_adapt_sp∂acesharing}(b) outlines the flow: the hardware is partitioned using the algorithm in this section, every subchip runs the full transpilation pipeline (routing, optimization, native decomposition), and the resulting circuits are merged into a single program for simultaneous execution.

Partitioning proceeds in three stages: seeding, growth, and rebalancing. During seeding circuit requests are handled from largest to smallest. The first region can select any available qubit; each subsequent region chooses the unassigned qubit that is farthest, according to the precomputed distance matrix, from all previously seeded qubits. This spreads demand across the chip and avoids early contention for the same high-quality hardware.

During the growth stage, all regions expand in lockstep. Each region maintains a priority queue of candidate qubits ordered by higher connectivity and lower noise penalty. The penalty is inspired from Das et al. which combines single-qubit error data with the average error on its incident couplers:
\[
\operatorname{penalty}(q)
    =   e^{(1)}_{q}
        + \frac{1}{\lvert \mathcal{N}(q) \rvert}
          \sum_{r \in \mathcal{N}(q)} e^{(2)}_{q,r}      
\]
where $e^{(1)}_{q}$ is the single-qubit error for qubit $q$, $e^{(2)}_{q,r}$ is the two-qubit error on edge $(q,r)$, $\mathcal{N}(q)$ denotes the neighbors of $q$ \cite{10.1145/3352460.3358287}. Regions first claim unassigned frontier qubits. If a region runs out of free neighbors, it may ``steal'' a leaf qubit from an adjacent region that still has surplus qubits. This helps keep regions connected without expensive connectivity checks.

\begin{table}[h]
\centering\normalsize
\caption{Compilation and fidelity metrics for circuits on \texttt{ibm\_brisbane}.}
\label{tab:mapomatic_results}

\begin{tabular}{|c|c|c|c|c|c|c|} 
\hline
Circuit & Qubits & $N_\mathrm{comp}$ & $t_\mathrm{comp}$ (ms) & $\bar{F}$ & $F_\mathrm{min}$ & $F_\mathrm{max}$ \\ 
\hline
bb84 & 8 & 1 & 6  & 0.938 & -- & -- \\ 
\hline
bb84 & 8 & 2 & 6  & 0.910 & 0.906 & 0.914 \\ 
\hline
bb84 & 8 & 4 & 5  & 0.892 & 0.766 & 0.942 \\ 
\hline
bb84 & 8 & 6 & 51 & 0.907 & 0.850 & 0.935 \\ 
\hline
qaoa & 6 & 1 & 39 & 0.600 & -- & -- \\ 
\hline
qaoa & 6 & 2 & 18 & 0.573 & 0.478 & 0.668 \\ 
\hline
qaoa & 6 & 4 & 25 & 0.556 & 0.383 & 0.682 \\ 
\hline
qaoa & 6 & 6 & 28 & 0.486 & 0.356 & 0.595 \\ 
\hline
qpe & 9 & 1 & 29 & 0.346 & -- & -- \\ 
\hline
qpe & 9 & 2 & 17 & 0.350 & 0.314 & 0.386 \\ 
\hline
qpe & 9 & 4 & 13 & 0.306 & 0.216 & 0.361 \\ 
\hline
qpe & 9 & 6 & 12 & 0.388 & 0.162 & 0.622 \\ 
\hline
\end{tabular}

\end{table}

In Table~\ref{tab:mapomatic_results} we test the partitioning algorithm described in this section and find promising results. We test three different circuits with increased conccurency on the chip and simulate the fidelity with DM-Sim from NWQ-Sim. For increased concurrent circuit execution we find that average circuit fidelity remains largely the same, with increased variability in each circuit.

%% file: Python.tex
\sloppy
\section{Python API} \label{sec:python}

QASMTrans provides a Python interface for programmatic access to transpilation and pulse control. We use \texttt{pybind11} to implement the interface. The \texttt{transpile\_file} function compiles OpenQASM programs against hardware specifications, supporting vendor-specific basis sets, pulse template injection, and configurable output formats. It returns transpilation artifacts including the mapped circuit, gate statistics, qubit assignments, pulse schedules, and critical-path metrics.

Pulse emission is handled through \texttt{prepare\_events} and \texttt{emit\_schedule}. The former generates time-ordered pulse events with channel assignments, amplitudes, durations, and phase tracking. The latter wraps event preparation to return execution summaries by default, or streams schedules directly to QICK hardware when enabled.

Device configuration is managed by \texttt{generate\_calibrated\_configs} and \texttt{build\_ankaa9q\_configs}. These functions construct device descriptions and pulse libraries from calibration data or devicelib metadata, with the latter supporting targeted qubit subsets and tunable waveform modeling parameters.

%% file: evaluation.tex
\begin{table}[h!]
\caption{Platforms for Portability Evaluation}
\centering\normalsize
\begin{tabular}{|c|c|c|c|c|c|} 
\hline
Platform                                               & CPU                                                            & Vendor & Core & Mem   & Compiler                                                     \\ 
\hline
\begin{tabular}[c]{@{}c@{}}MacBook \\Pro\end{tabular} & Apple M2                                                             & Apple  & 12   & 16GB  & \begin{tabular}[c]{@{}c@{}}AppleClang\\~14.0.3\end{tabular}  \\ 
\hline
Perlmutter                                             & \begin{tabular}[c]{@{}c@{}}Authentic\\AMD\end{tabular}         & AMD    & 128  & 256GB & g++ 11.2.0                                                   \\ 
\hline
JetsonTX2                                              & ARMV8                                                          & NVIDIA & 4    & 8GB   & g++ 5.4.0                                                    \\ 
\hline
Crusher                                                & \begin{tabular}[c]{@{}c@{}}Authentic\\AMD\end{tabular}         & AMD    & 128  & 512GB & g++ 12.2.0                                                   \\ 
\hline
Frontier                                               & \begin{tabular}[c]{@{}c@{}}Authentic\\AMD\end{tabular}         & AMD    & 128  & 512GB & g++ 12.2.0                                                   \\ 
\hline
Summit                                                 & POWER9                                                         & IBM    & 176  & 512GB & g++ 9.1.0                                                    \\ 
\hline
Tonga                                                  & Intel P8168                                                    & Intel  & 96   & 128GB & g++ 11.2.0                                                   \\ 
\hline
Theta                                                  & \begin{tabular}[c]{@{}c@{}}Intel Phi \\7230 (KNL)\end{tabular} & Intel  & 256  & 192GB & intel 19.1.0                                                 \\
\hline
\end{tabular}
\label{tab:platform}
\end{table}

\section{Evaluation} \label{sec:evaluation}

\subsection{Experimental setup}

We primarily use the NERSC Perlmutter HPC system for our evaluation. Perlmutter is built by HPE. Each of the Cray EX systems is equipped with an AMD EPYC 7763 CPU and four NVIDIA A100 GPUs. The other platforms used for the transpilation are listed in Table~\ref{tab:platform}. We compare QASMTrans to the  Qiskit transpiler which uses the sabre algorithm algorithm~\cite{li+:asplos19,IBMQiskit}. We focus on transpilation efficiency, quality, and fidelity. The efficiency is measured by transpilation time. The quality is measured by the depth, total number of gates, and number of \texttt{CX} gates of the transpiled circuit. The fidelity is measured by calculating the fidelity of execution for the transpiled circuit over four real quantum devices: \emph{IBM-Brisbane}, \emph{Rigetti-AspenM2}, \emph{IonQ-Aria1} and \emph{Quantinuum-H1-1}). We test on different benchmark circuits varying from 10 qubits to 127 qubits from QASMBench~\cite{li2023qasmbench}, We show all the benchmark information in Table ~\ref{tab:benchmark_information}.

\subsection{Transpilation Efficiency and Quality}

The evaluation results are listed in Table~\ref{tab:Compilation_time_analysis}. We use IBMQ devices as the transpilation target so that: (i) the basis gate set is \texttt{X}, \texttt{SX}, \texttt{CX}, and \texttt{RZ}; (ii) for topology, when the number of qubits of the circuit is less than 27, we use the topology of IBMQ Toronto. When it is larger than 27, we use the topology of the latest 127-qubit IBM Brisbane as the objective device. 

\begin{table}[h]
\centering\normalsize
\caption{Benchmark information, it shows the qubit number, single-qubit and total gates, and the depth of the circuit}
\label{tab:benchmark_information}

\begin{tabular}{|c|c|c|c|c|} 
\hline
\textbf{Benchmarks}        & \multicolumn{4}{c|}{\textbf{Circuit Information}}  \\ 
\hline
Name              & Qubits & 1-q gate & total gate & Depth    \\ 
\hline
square\_root & 18     & 1415     & 2313       & 1269     \\ 
\hline
vqe\_uccsd     & 8      & 5320     & 10K        & 7252     \\ 
\hline
sat          & 11     & 53       & 53         & 51       \\ 
\hline
bwt          & 21     & 66K      & 87K        & 53K      \\ 
\hline
gcm           & 13     & 2387     & 3149       & 2447     \\ 
\hline
hhl          & 7     & 493     & 689      & 551     \\ 
\hline
qaoa          & 6      & 222      & 276        & 110      \\ 
\hline
qec               & 5      & 20       & 30         & 18       \\ 
\hline
adder             & 4      & 17       & 27         & 12       \\ 
\hline
adder             & 10     & 10       & 35         & 24       \\ 
\hline
adder             & 64     & 93       & 212        & 78       \\ 
\hline
adder & 118 & 2861 & 3706 & 132 \\
\hline
qpe   & 9 & 15 & 31 & 21 \\
\hline
bb84 & 8 & 27 & 27 & 7 \\
\hline
bv & 14 & 28 & 41 & 17 \\
\hline
qugan & 111 & 2039 & 2697 & 112 \\
\hline
qv & 32 & 6144 & 7680 & 320 \\
\hline
qram & 20 & 545 & 681 & 24 \\
\hline
\end{tabular}
\end{table}

\vspace{4pt}\noindent\textbf{Quality:} Overall, QASMTrans can generate transpiled circuits with comparable depth, gates and 2-qubit gates as Qiskit. The observed difference is from currently un-implemented gate transformation, cancellation passes, as well as heavier routing techniques.  

\vspace{4pt}\noindent\textbf{Efficiency:} As listed, QASMTrans shows a large performance advantage over Qiskit for the 16 benchmark circuits. The speedup can be as much as 171$\times$ over Qiskit. In particular, for some challenging circuits, such as the bwt\_n21 with 87K gates QASMTrans transpiles in 4$\times$ the speed with similar circuit quality.

\vspace{4pt}\noindent\textbf{Scalability:} We further look at the performance scalability. Figure~\ref{fig:compilation_time} shows the scaling of the transpilation time with respect to the number of gates of the input circuits for the various benchmarks. As can be seen, the performance advantage over Qiskit is quite consistent.

\subsection{Transpilation Fidelity}

\begin{figure}[h!]
  \centering
  \includegraphics[width=0.5\linewidth]{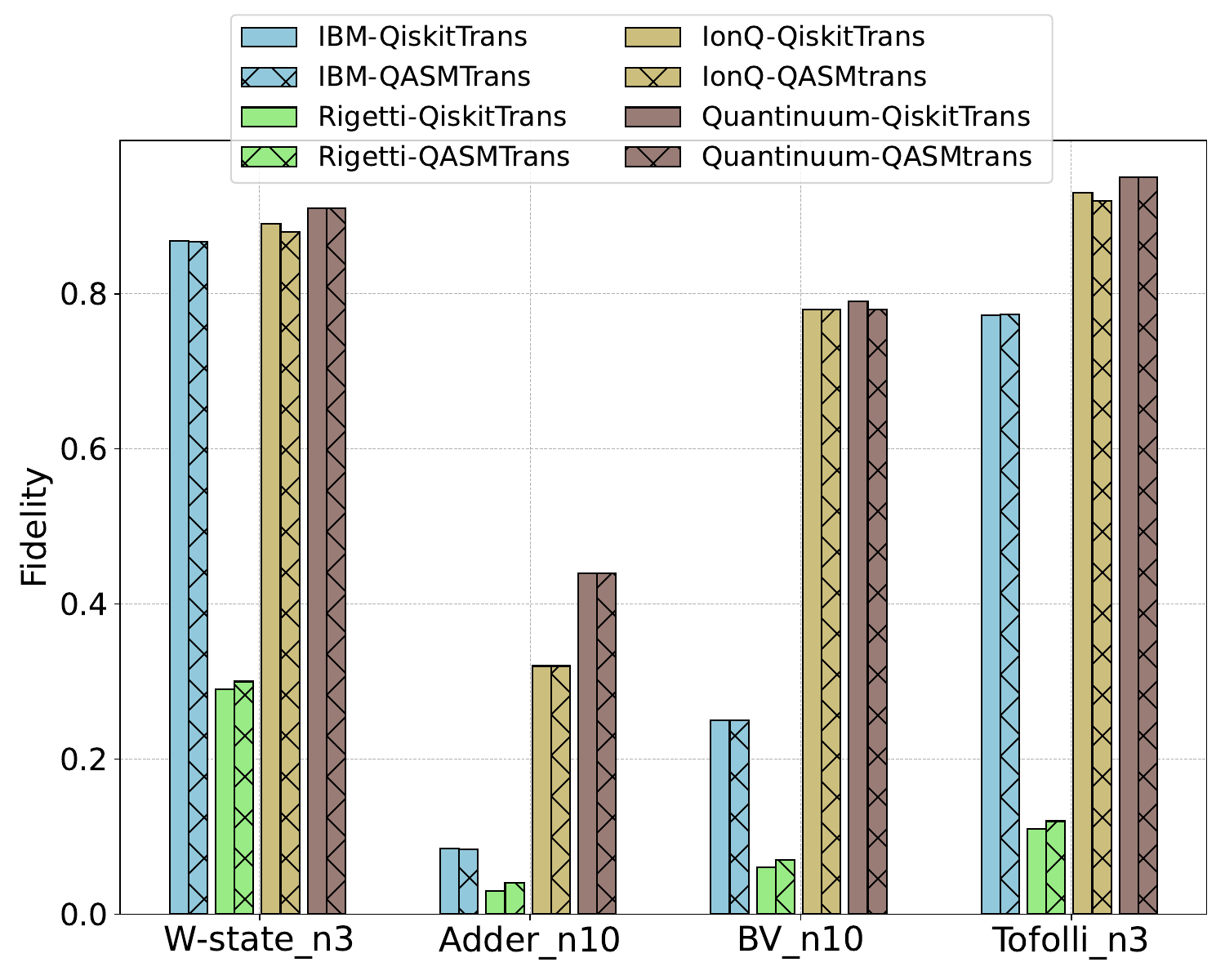}
  \caption{QASMTrans fidelity analysis compared with Qiskit transpiler on different machines, the X-axis shows the benchmarks, and Y-axis shows the fidelity obtained on real NISQ machines (IBM \emph{ibm\_brisbane}, Rigetti \emph{Aspen-M2}, IonQ \emph{Aria-1}, and Quantinuum \emph{H2}) with respect to the Qiskit results.}
  \label{fig:fidelity}
\end{figure}

To evaluate the correctness of transpilation, we use the transpiled circuits generated by Qiskit and QASMTrans as the inputs, and launch them onto four real NISQ devices (IBMQ, Rigetti, Quantinuum, and IonQ) to assess the difference in their induction results, shown in Figure~\ref{fig:fidelity}. Please be aware that these input circuits, despite having already been transpiled, may go through another round of internal transpilation or optimization within the backend processing of the NISQ device. This is not under our control. However, we argue that this will not significantly impact the fidelity results since both input circuits go through the same backend processes.

As can be seen in Figure~\ref{fig:fidelity}, the quantum state fidelity with Qiskit result is quite consistent across input circuits and underlying hardware, with $<1\%$ deviation.

\subsection{Optimization for Classical Simulation}

Both the constrained qubit routing/mapping and user-guided qubit prioritization presented in Section \ref{sec:transpiler} can harvest performance gain for classical simulation. Constrained qubit routing/mapping limits the number of qubits for the simulation, for which the performance gain is quite obvious. Here, we mainly focus on demonstrating the benefit of user-guided qubit prioritization.

We have already discussed why minimizing the number of gates over the global qubits can reduce the overhead from communication. Here, we use SV-Sim~\cite{li2021sv} as the classical simulator. We use all the 8 GPUs from 2 Perlmutter nodes for the distributive circuit simulation. Consequently, 3 qubits are sharing their corresponding coefficients across the 8 GPUs. Figure~\ref{fig:remap} shows the difference in simulation time for the transpiled circuits with and without user-guided qubit prioritization. The performance gain can be quite significant given the log-scale of the Y-axis. This benefit mainly comes from switching some expensive gates over the three global qubits to local qubits through the final remapping of qubit prioritization.

\subsection{Platform Portability}

\begin{table*}[b]
\centering\small
\caption{Evaluation of QASMTrans compared to Qiskit in terms of transpilation quality and efficiency. QT represents our QASMTrans method, O1/O2/O3 represent three optimization levels for Qiskit.}
\label{tab:Compilation_time_analysis}

\begin{tabular}{|c|c|c|c|c|c|c|l|l|l|} 
\hline
\multicolumn{2}{|c|}{Benchmarks} & \multicolumn{5}{c|}{Efficiency: Transpilation Time (ms)} & \multicolumn{3}{c|}{Quality: Transpiled by Qiskit-O1/O2/O3/QASMTrans} \\ 
\hline
Name & \# qubit & O1 & O2 & O3 & QT & O1/QT & \multicolumn{1}{c|}{1Q Gate} & \multicolumn{1}{c|}{2Q Gate} & \multicolumn{1}{c|}{Depth} \\ 
\hline
qv & 32 & 4.6K & 348 & 398 & 55 & 85 & 30.6K/19.4K/19.3K/20.5K & 5.1K/4.9K/4.9K/7.3K & 4.2K/3.3K/3.4K/5.6K \\ 
\hline
adder & 4 & 6 & 6 & 7 & 0.07 & 88 & 17/16/16/21 & 13/13/13/19 & 20/19/19/28 \\ 
\hline
adder & 10 & 9 & 20 & 36 & 0.32 & 29 & 101/146/143/141 & 128/106/106/128 & 194/196/199/211 \\ 
\hline
adder & 64 & 46 & 113 & 136 & 11 & 4 & 5.9K/3.8K/3.6K/0.98K & 1.2K/1.0K/1.0K/2.4K & 2.0K/1.4K/1.4K/1.3K \\ 
\hline
adder & 118 & 79 & 185 & 276 & 48 & 2 & 9.6K/7.3K/6.6K/1.8K & 2.0K/2.0K/1.8K/6.2K & 3.0K/2.7K/2.5K/2.4K \\ 
\hline
bv & 14 & 8 & 8 & 8 & 0.07 & 112 & 33/34/34/106 & 11/9/9/11 & 23/22/22/23 \\ 
\hline
bwt & 21 & 5,200 & 25,600 & 35,000 & 1,430 & 4 & 277K/494K/491K/511K & 585K/512K/517K/630K & 487K/513K/518K/558K \\ 
\hline
gcm & 13 & 17 & 96 & 108 & 4 & 5 & 2.4K/2.3K/2.3K/2.4K & 1.2K/0.8K/0.8K/1.4K & 2.9K/2.0K/2.0K/3.0K \\ 
\hline
hhl & 7 & 7 & 38 & 30 & 1 & 6 & 794/351/334/794 & 365/183/182/430 & 912/367/367/984 \\ 
\hline
qaoa & 6 & 9 & 12 & 12 & 0.52 & 17 & 324/200/214/1224 & 81/60/64/96 & 214/116/134/579 \\ 
\hline
qec & 5 & 7 & 7 & 9 & 0.04 & 171 & 4/4/4/1 & 4/4/4/10 & 10/10/10/11 \\ 
\hline
qram & 20 & 9 & 27 & 19 & 0.66 & 14 & 138/258/253/243 & 252/200/200/248 & 240/240/243/297 \\ 
\hline
qugan & 111 & 83 & 214 & 232 & 33 & 3 & 9.8K/6.5K/6.2K/7.2K & 2.0K/1.6K/1.5K/4.4K & 3.6K/2.2K/2.2K/3.5K \\ 
\hline
sat & 11 & 14 & 37 & 47 & 1 & 10 & 479/791/845/823 & 594/481/481/615 & 756/741/836/868 \\ 
\hline
square\_root & 18 & 35 & 124 & 183 & 6 & 6 & 1.5K/2.4K/2.5K/2.8K & 2.4K/2.0K/2.0K/2.7K & 2.4K/2.6K/2.6K/3.1K \\ 
\hline
vqe\_uccsd\ & 8 & 63 & 176 & 285 & 16 & 4 & 4.4K/7.5K/7.4K/24.1K & 5.8K/5.1K/5.1K/7.0K & 7.9K/8.0K/8.0K/14.9K \\ 
\hline
\end{tabular}
\end{table*}

We evaluate QASMTrans across different computing platforms, from various HPC systems, including NERSC Perlmutter, OLCF Frontier, Crusher, and Summit, ALCF Theta, to a desktop and laptop (Intel P8168 and Apple M2), to an embedded device (JetsonTX2 with ARM8). The platforms are listed in Table~\ref{tab:platform}. The results are shown in Figure~\ref{fig:fidelity_platform}. The transpilation on all the platforms can be finished within 100 seconds and most of them below 1 second. 

With these results, we have three observations: (i) QASMTrans can be portable on various platforms, given its efficient C++ based implementation and non-external library dependency (the \emph{json} and \emph{lexertk} are included as header files). In particular, the successful and efficient running on an ARM8 CPU shows the potential of practical deployment on an FPGA of a real quantum system or testbed, such as LBNL AQT. (ii) The transpilation speed across applications circuits and platforms is consistent. (iii) The majority (nearly $90\%$) of the transpilation time is devoted to routing and mapping for the current implementation of QASMTrans.  

%% file: relatedwork.tex
\section{Related Work}

\subsection{Quantum Intermediate Representation}

In quantum computing, gate IR provides an essential abstraction layer, offering a structured, machine-agnostic representation of quantum circuits. Among the existing quantum IRs, the Microsoft QIR~\cite{qir} is an LLVM-based IR that defines a set of rules for representing quantum constructs. QIR attempts to serve as a common interface between various quantum languages (e.g., Q\#) and platforms. QASM~\cite{cross2017open} is a widely recognized quantum assembly language developed by IBM for its hardware platforms and software tool-chain. Quil is a portable quantum instruction language developed by Rigetti. Lastly, XACC (eXtreme-scale ACCelerator)~\cite{mccaskey2020xacc} is a compilation framework for hybrid quantum-classical computing architectures developed at ORNL, supporting IBM, Rigetti, D-Wave QPUs, and various classical simulators such as SV-Sim~\cite{li2021sv} and DM-Sim~\cite{li2020density}.

\subsection{Quantum Transpilation}

Quantum transpiler plays a crucial role in quantum computing by translating high-level quantum algorithms into a series of low-level hardware-specific instructions that quantum hardware can execute.
Qiskit is a widely used quantum software development package developed by IBM. The Qiskit transpiler provides a flexible and extensible framework, offering a wide array of compilation passes that can be combined in different ways to create customized and hardware-tailored transpilation pipelines. 

In addition to Qiskit, there are various transpilers aiming at different purposes:
1) \emph{application-oriented transpilation:} These transpilers focus on specific domain applications. For example, Paulihedral~\cite{li+:asplos22} focuses on VQE, Twoqan~\cite{lao+:arxiv21twoqan} concentrates on QAOA circuits.
2) \emph{hardware-oriented transpilation:} These transpilers focus on supporting the new features of a particular quantum platform. For instance, CaQR emphasizes the support for dynamic circuit generation and the opportunities from qubit reset~\cite{hua2023caqr}. Pulse transpilers delve into the nuances of low-level pulse scheduling, optimizing quantum operations at the physical layer \cite{gokhale2020optimized, chen2023pulse, shi+:asplos19}. AutoComm~\cite{wu2022autocomm} and QuComm~\cite{wu2022collcomm} present transpiler optimization techniques for distributive quantum devices.
3) \emph{Optimization for mapping/routing:} there are also works aimed at improving general transpilation performance, like Sabre~\cite{li+:asplos19} and Zulhner ~\cite{zulehner+:date18:} attempt to minimize the number of additional gates in mapping/routing. TOQM \cite{zhang+:asplos21} aims at shrinking the depth of the transpiled circuit. Shi {\it et al.}~\cite{shi+:asplos19} presents the complete transpilation and optimization flow, including gate aggregation and cancellation. QASMTrans falls into the third category, aiming at improving the transpilation performance of large and deep QASM circuits.

\begin{figure}[h!]
  \centering
  \includegraphics[width=0.55\linewidth]{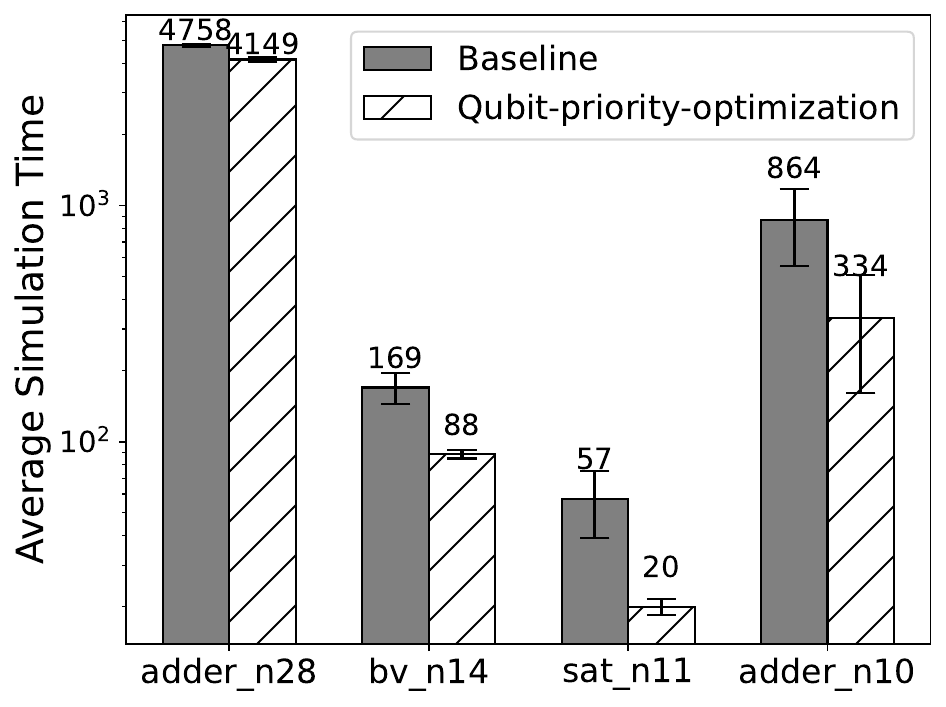}
  \caption{Performance gain in classical simulation through user-guided qubit prioritization using SV-Sim for the transpiled circuits on 8 GPUs of Perlmutter. Note, the Y-axis of simulation time is in log-scale.}
  \label{fig:remap}
\end{figure}

\subsection{Pulse-Level Control and Optimization}
Pulse-level control represents the lowest level of quantum program compilation, translating gate-level operations into time-domain waveforms that directly drive quantum hardware. Several frameworks have explored this space with varying degrees of device specificity and optimization capability.

Qiskit Pulse~\cite{alexander+:qiskitpulse20} pioneered accessible pulse-level programming by extending the Qiskit framework with pulse scheduling primitives. However, its integration is limited to IBM hardware and lacks support for portable pulse generation across different vendor platforms. OpenPulse provides a standardized pulse IR within OpenQASM 3.0, enabling pulse-level descriptions but requiring vendor-specific backends for execution ~\cite{capelluto2020openpulse}. XACC offers pulse compilation capabilities but lacks comprehensive device-specific pulse libraries and calibration-aware optimization~\cite{9347736}.

Device-specific pulse control frameworks include Pulser for neutral atom systems, JaqalPaw for trapped-ion devices, and eQASM which provides near-binary pulse instruction representation for superconducting systems~\cite{Silverio2022pulseropensource,lobser2023jaqalpawguidedefiningpulses,fu2019eqasmexecutablequantuminstruction}. While these frameworks offer fine-grained control, their tight coupling to specific hardware platforms limits portability and cross-platform experimentation.

Recent work has explored pulse-level optimization to improve circuit execution. Wei et al. demonstrated building arbitrary native two-qubit gates on IBM systems beyond standard cross-resonance interactions ~\cite{PRXQuantum.5.020338}. Gokhale et al.  introduced pulse-level compilation for near-term algorithms using OpenPulse ~\cite{capelluto2020openpulse}. Chen et al. comprehensively studied mining transpiled circuits for frequent sub-circuits and merging gates to decrease latency through pulse-level synthesis ~\cite{chen2023pulse}. Their work focuses on discovering merge candidates through exhaustive circuit analysis, whereas QASMTrans takes a complementary approach by targeting the critical path and designing customized pulse gates for frequently used logical operations, reducing discovery overhead while enabling substantial latency reductions.

\begin{figure*}[h!]
  \centering
  \includegraphics[width=0.95\linewidth]{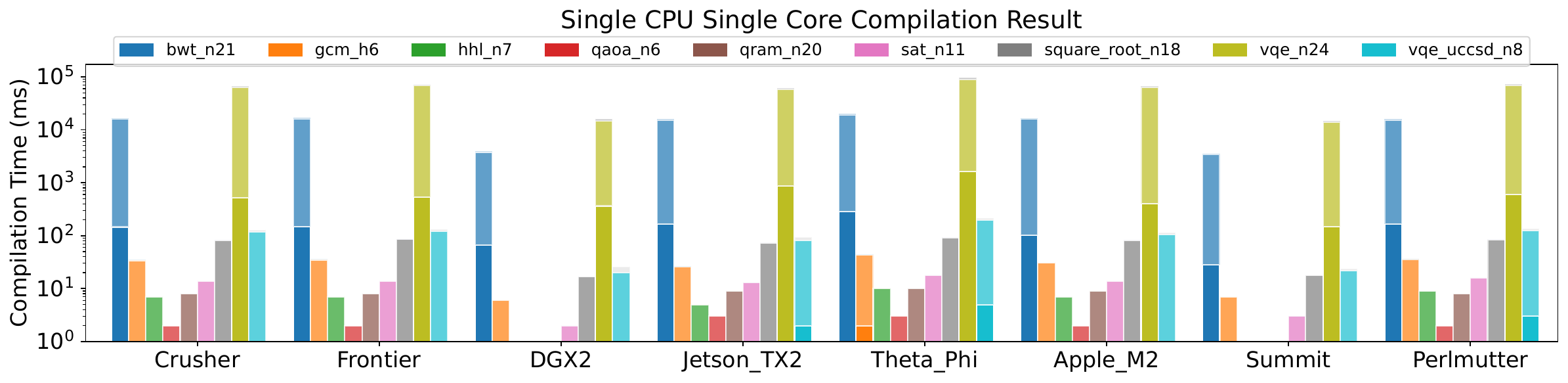}
  \caption{Compilation time on various platforms. The X-axis shows the names of different platforms, while the Y-axis is the compilation time using a single core of a CPU of the system. The empty bars indicate the condition that the compilation time is less than 1ms. The breakdown of each bar implies the time of (upper) routing \& mapping, and (lower) decomposition. Please be aware that the Y-axis is in the log scale.
 }
  \label{fig:fidelity_platform}
\end{figure*}\label{sec:related_work}

Control system integration represents another critical dimension. QICK provides an open-source FPGA-based control platform for rapid prototyping and low-latency adaptive control~\cite{Stefanazzi_2022}. QuBiC offers comprehensive calibration and benchmarking workflows for superconducting qubits~\cite{xu_qubic_2021}. QASMTrans distinguishes itself by providing end-to-end compilation from QASM to calibrated pulse schedules with direct QICK integration for embedded deployment, validated through QuTiP-based pulse-level simulation. This integration enables just-in-time compilation for adaptive algorithms while maintaining device portability through standardized configuration files.